\documentclass[journal,comsoc]{IEEEtran}
\usepackage[T1]{fontenc}
\usepackage{cite, color}
\usepackage[normalem]{ulem}
\usepackage{longtable}
\usepackage{float}
\usepackage[acronym,nonumberlist,nogroupskip,style=super]{glossaries}
\usepackage{array}
\usepackage{multirow}
\usepackage{booktabs}
\usepackage{multirow}
\ifCLASSINFOpdf
   \usepackage [pdftex]{graphicx}
   \DeclareGraphicsExtensions{.pdf,.jpg,.png,.eps}
\else
	\usepackage [dvips]{graphicx}
	\DeclareGraphicsExtensions{.eps}
\fi
\usepackage{amsmath}
\usepackage{inputenc}
\usepackage{amssymb}
\usepackage{enumitem}
\usepackage[cmintegrals]{newtxmath}
\ifCLASSOPTIONcompsoc
  \usepackage[caption=false,font=normalsize,labelfont=sf,textfont=sf]{subfig}
\else
  \usepackage[caption=false,font=footnotesize]{subfig}
\fi
\newcommand{\rulesep}{\unskip\ \vrule\ }

\newacronym{e2e}{E2E}{end-to-end}
\newacronym{m2m}{M2M}{Machine-to-Machine}
\newacronym{iot}{IoT}{Internet of Things}
\newacronym{rpl}{RPL}{Routing Protocol for Low Power and Lossy Networks}
\newacronym{ids}{IDS}{Intrusion Detection System}
\newacronym{rfid}{RFID}{Radio Frequency Identification}
\newacronym{wsn}{WSN}{Wireless Sensor Network}
\newacronym{6lowpan}{6LoWPAN}{IPv6 over Low-powered Wireless Personal Area Network}
\newacronym{ietf}{IETF}{Internet Engineering Task Force}
\newacronym{ipv6}{IPv6}{Internet Protocol version 6}
\newacronym{tcp}{TCP}{Transmission Control Protocol}
\newacronym{of}{OF}{Objective Function}
\newacronym{dag}{DAG}{Directed Acyclic Graph}
\newacronym{dodag}{DODAG}{Destination Oriented Directed Acyclic Graph}
\newacronym{ch}{CH}{Cluster Head}
\newacronym{roll}{ROLL}{Routing Over Low-power and Lossy Networks working group}
\newacronym{of0}{OF0}{Objective Function Zero}
\newacronym{mrhof}{MRHOF}{Minimum Rank with Hysteresis Objective Function}
\newacronym{ext}{EXT}{Expected Transmission Count}
\newacronym{extof}{EXTOF}{Expected Transmission Count Objective Function}
\newacronym{lbof}{LBOF}{Load Balancing Objective Function}
\newacronym{cnc}{CNC}{Child Node Count}
\newacronym{taof}{TAOF}{Traffic Aware Objective Function}
\newacronym{ptr}{PTR}{Packet Transmission Rate}
\newacronym{dodagid}{DODAGID}{DODAG Identification}
\newacronym{dio}{DIO}{DODAG Information Object}
\newacronym{dis}{DIS}{DODAG Information Solicitation}
\newacronym{dao}{DAO}{Destination Advertisement Object}
\newacronym{daoack}{DAO-ACK}{DAO Acknowledgement}
\newacronym{P2P}{P2P}{Point to Point communication}
\newacronym{P2MP}{P2MP}{Point to Multi-Point communication}
\newacronym{MP2P}{MP2P}{Multi-Point to Point communication}
\newacronym{cc}{CC}{Consistency Check}
\newacronym{icmp}{ICMPv6}{Internet Control Message Protocol}
\newacronym{ccm}{CCM}{Counter with CBC-MAC "Cipher Block Chaining - Message Authentication Code"}
\newacronym{aes}{AES}{Advanced Encryption Standard}
\newacronym{macl}{MAC}{Medium Access Control}
\newacronym{macc}{MAC}{Message Authentication Code}
\newacronym{rsa}{RSA}{Rivest-Shamir-Adleman encryption}
\newacronym{sha}{SHA}{Secure Hash Algorithm}
\newacronym{os}{OS}{Operating System}
\newacronym{gps}{GPS}{Global Positioning System}
\newacronym{rtt}{RTT}{Round-Trip Time}
\newacronym{phy}{PHY}{Physical layer}
\newacronym{coap}{CoAP}{Constrained Application Protocol}
\newacronym{ipsec}{IPSec}{Internet Protocol Security}
\newacronym{dtls}{DTLS}{Datagram Transport Layer Security protocol}
\newacronym{rssi}{RSSI}{Received Signal Stregnth Indicator}
\newacronym{rss}{RSS}{Received Signal Stregnth}
\newacronym{dos}{DoS}{Denial of Service}
\newacronym{esp}{ESP}{Encapsulated Security Header}
\newacronym{tpm}{TPM}{Trusted Platform Module}
\newacronym{ernt}{ERNT}{Extended RPL Node Trustworthiness}
\newacronym{tof}{TOF}{Trust Objective Function}
\newacronym{srpl}{SRPL}{Secure RPL}
\newacronym{dht}{DHT}{Distributed Hash Table}
\newacronym{trail}{TRAIL}{Trust Anchor Interconnection Loop}
\newacronym{vera}{VeRA}{Version attack and Rank Authentication}
\newacronym{mop}{MOP}{Mode of Operation}
\newacronym{mop2}{MOP2}{mode of operation}
\newacronym{sprt}{SPRT}{Sequential Probability Ratio Test}
\newacronym{qos}{QoS}{Quality of Service}
\newacronym{ami}{AMI}{Advanced Metering Infrastructure}
\newacronym{cr}{CR}{Cognitive Radio}
\newacronym{loadng}{LOADng}{Low power and Lossy Networks On-demand Ad-hoc Distance-vector routing protocol - Next Generation}
\newacronym{loadng-ctp}{LOADng-CTP}{LOADng with Collection Tree Protocol}
\newacronym{corpl}{CORPL}{Cognitive and Opportunistic RPL}
\newacronym{aodv}{AODV}{Ad-hoc On-demand Distance Vector}
\newacronym{pdr}{PDR}{packet delivery rate}
\newacronym{rdc}{RDC}{Radio Duty-Cycle}
\newacronym{umrpl}{UM}{Unsecured Mode}
\newacronym{psmrpl}{PSM}{Preinstalled Secure Mode}
\newacronym{asmrpl}{ASM}{Authenticated Secure Mode}

\makeglossaries
\glsaddall
\IEEEaftertitletext{\vspace{-1\baselineskip}}

\begin{document}
%
\title{Enhancing Routing Security in IoT: Performance Evaluation of RPL's Secure Mode under Attacks}

\author{Ahmed~Raoof,~\IEEEmembership{Student~Member,~IEEE,}
	Ashraf~Matrawy,~\IEEEmembership{Senior~Member,~IEEE,}
	and~Chung-Horng~Lung,~\IEEEmembership{Senior~Member,~IEEE}
	\thanks{A. Raoof and C. Lung are with the Department of Systems and Computer Engineering,
		Faculty of Engineering and Design, Carleton University, Ottawa, ON, Canada (email: ahmed.raoof@carleton.ca; chlung@sce.carleton.ca)}
	\thanks{A. Matrawy is with the School of Information Technology, Carleton University, Ottawa, ON, Canada (email: ashraf.matrawy@carleton.ca)}
}

\maketitle

\markboth{IEEE~Internet of Things,~Vol.~xx, No.~x, XXX~2020}{~Raoof \MakeLowercase{\textit{et al.}}: Enhancing Routing Security in IoT: Performance Evaluation of RPL’s Secure Mode under Attacks}

\begin{abstract}
As the \gls{rpl} became the standard for routing in the \gls{iot} networks, many researchers had investigated the security aspects of this protocol. However, no work (to the best of our knowledge) has investigated the use of the security mechanisms included in \gls{rpl}'s standard, mainly because there was no implementation for these features in any \gls{iot} operating systems yet. A partial implementation of \gls{rpl}'s security mechanisms was presented recently for the Contiki operating system (by Perazzo \textit{et al.}), which provided us with an opportunity to examine \gls{rpl}'s security mechanisms. In this paper, we investigate the effects and challenges of using \gls{rpl}'s security mechanisms under common routing attacks. First, a comparison of \gls{rpl}'s performance, with and without its security mechanisms, under four routing attacks (Blackhole, Selective-Forward, Neighbor, and Wormhole attacks) is conducted using several metrics (e.g., average data packet delivery rate, average data packet latency, average power consumption, etc.). This comparison is performed using two commonly used Radio Duty-Cycle protocols. Secondly, and based on the observations from this comparison, we propose two techniques that could reduce the effects of such attacks, without having added security mechanisms for \gls{rpl}. An evaluation of these techniques shows improved performance of \gls{rpl} under the investigated attacks, except for the Wormhole attack.
\end{abstract}


\section{Introduction}
Routing is one of the most researched fields in the world of \gls{iot}, due to the constraint nature of these devices. Introduced by \gls{ietf}, \gls{rpl} \cite{RFC6550} had become the standard for routing in many \gls{iot} networks as it was designed to efficiently use the constrained resources of \gls{iot} devices, while providing effective routing services. Routing security was an integral part of \gls{rpl}'s design with several, but optional, security mechanisms available\cite{RFC6550}.

Since it became a standard in 2012, \gls{rpl} gained a great deal of research interest, with many of the literature focusing on the security aspects of routing using the protocol, such as types of routing attacks, new mitigation methods and \glspl{ids}, and security-minded \glspl{of}\cite{TAOF,Djedjig2015,Djedjig2017, Karkazis2014,Airehrour2019}. Interestingly, there has been no research discussing the effects of using \gls{rpl}'s security mechanisms, specifically under routing attacks. This is most probably due to the lack of implementation of \gls{rpl}'s security mechanisms in any of the available \gls{iot} \glspl{os}, such as Contiki \gls{os}\cite{ContikiOSRef} and TinyOS \cite{TinyOS}.

However, recently Perazzo \textit{et al.} in \cite{Perazzo2017} provided a partial implementation of \gls{rpl}'s security mechanisms for Contiki OS, which added \gls{psmrpl} and the optional replay protection mechanism. This implementation provided us with the basis upon which the work in this paper is built on. In this paper, we have experimentally investigated \gls{rpl}'s performance under four common routing attacks using several metrics to analyze and compare the performance between having \gls{rpl}'s security mechanisms enabled or disabled.

The work in this paper provides a significant extension to our previous conference paper\cite{Raoof2019a}. Specifically, we first introduced a new scenario for an \gls{rpl} Wormhole attack to the evaluation. Then, we extended the evaluation of \gls{rpl}'s performance from using one \gls{rdc} protocol (the ContikiMAC protocol) to include the effect of using another commonly used \gls{rdc} protocol, namely the NullRDC protocol - see \S\ref{ImptCh}. Finally, for the two techniques we proposed in \cite{Raoof2019a} to improve \gls{rpl}'s performance under the investigated attack, we conducted an extensive evaluation of these two techniques and their effects on \gls{rpl}'s performance under the investigated attacks and using the two underlying \gls{rdc} protocols.

Our contributions can be summarized as follows:
\begin{itemize}
	\item Through more than a thousand experiments, we provided a performance comparison for \gls{rpl} between the \gls{umrpl} and \gls{psmrpl}; the latter is examined with and without the optional replay protection. We showed that running \gls{rpl} in \gls{psmrpl} (without replay protection) does not use more resources than \gls{umrpl}, even under an attack.
	\item We verified that \gls{rpl} in \gls{psmrpl} can stop external adversaries from joining the \gls{iot} network for the investigated attacks, except for the Wormhole attack. Furthermore, we showed that the optional replay protection also provides excellent mitigation against the Neighbor attack. However, it needs further optimization to reduce its effect on energy consumption.
	\item We observed and analyzed the effect of the investigated attacks on the routing topology and proposed two simple techniques that could help reduce the effects of the investigated attacks, without using external security measures such as IDSs or added security mechanisms.
	\item Another performance comparison of the implementation of the proposed techniques was conducted. The results showed improved performance of \gls{rpl} under the Blackhole and Selective-Forward attacks, in terms of \gls{pdr} and \gls{e2e} latency.
\end{itemize}

The rest of this paper goes as follows: Section \ref{rltdwrk} looks into the related works. In section \ref{background} an overview of \gls{rpl} and its security mechanisms is presented. Section \ref{RPLEval} discusses our evaluation methodology, setup, assumptions, adversary model, and attack scenarios. Evaluation results are analyzed in section \ref{resultsanalysis}. Section \ref{discussionSec} discuses our observations from the results and proposes two suggestions to be used when designing \gls{rpl}-based \gls{iot} networks. In addition, an implementation of the proposed suggestions is evaluated and the results are discussed. Finally, our work is concluded in \ref{Conc}.

\section{Related Works}\label{rltdwrk}
This section highlights some influencing literature that discussed \gls{rpl}'s performance under common routing attacks. As stated earlier, none of them had investigated \gls{rpl}'s security mechanisms, except for the conference version of this paper.

Le \textit{et al.} in \cite{Le2013a} evaluated \gls{rpl}'s performance under four \gls{rpl}-based attacks: the Decreased Rank, Local Repair, Neighbor, and DODAG\footnote{DODAG = Destination-Oriented Directed Acyclic Graph} Information Solicitation (DIS) attacks. Their work showed that the Decreased Rank and the Local Repair attacks affect the \gls{pdr} the most, while the \acrshort{dis} attack introduced the most \gls{e2e} latency. The Neighbor attack showed the least impact on the network. Compared to our work, the authors only tackled with the unsecured mode of \gls{rpl} while ignoring the effect of their attacks on power consumption.

Kumar \textit{et al.} in \cite{Kumar2016} investigated the effects of the Blackhole attack, on \gls{rpl}-based network through simulations. As expected, the attack was successful in reducing the \gls{pdr} and increased both the \gls{e2e} latency and control messages overhead. However, the authors did not evaluate the power consumption and neglected the existence of \gls{rpl}'s security mechanisms.

Perazzo \textit{et al.} in \cite{Perazzo2017, Arena2020} provided the first, standard-compliant as per their claim, partial implementation of \gls{rpl} security mechanisms. One secure mode, the Preinstalled secure mode, and the optional replay protection, named the \gls{cc} mechanism, were introduced to ContikiRPL (Contiki OS version of \gls{rpl}). The authors provided an evaluation for their implementation and compared \gls{rpl}'s performance between using and not using the \gls{psmrpl}. However, It is worth noting that the authors did not evaluate their implementation against actual attacks.

Our previous work in \cite{Raoof2019} presented the first glimpse of the effects that \gls{rpl}'s security mechanisms could have on \gls{rpl}-based \gls{iot} networks when there is an actual attack. \gls{rpl}'s performance (with and without \gls{psmrpl}) was investigated under three attacks: the Blackhole, Selective-Forward, and Neighbor attacks using simulations. The preliminary results showed that \gls{rpl}'s secure modes can mitigate the external adversaries of the investigated attacks, not the internal ones. However, it did not provide an in-depth analysis of the results nor inspected the optional replay protection mechanism.

\section{Background Review}\label{background}

\subsection{RPL Overview} \label{RPLOverview}
\gls{rpl} was developed as a distance-vector routing protocol\cite{RFC6550}. It arranges the network devices into a \glspl{dodag}\cite{Janicijevic2011}: a network of nodes connected without loops and where the traffic is directed toward one \textit{root} or \textit{sink} node\cite{RFC6550,Granjal}.

The creation of the \gls{dodag} depends on the used \textit{\gls{of}}, which defines essential configurations such as the used routing metrics, how to calculate the \textit{rank}\footnote{The rank of a node represents its distance to the root node based on the routing metrics defined by the \gls{of}}, and how to select parents in the \gls{dodag}. To accommodate the different applications and environments where \gls{rpl} can be deployed, \gls{rpl} has several \glspl{of}\cite{RFC6552, RFC6719, TAOF} available for use\cite{Raoof2018}. Also, deployments of \gls{rpl} can have their own \glspl{of}.

\gls{rpl} supports three types of traffic: \gls{MP2P} traffic (nodes to sink) through normal \gls{dodag}, \gls{P2MP} traffic (sink to nodes) through source routing, and \gls{P2P} traffic (non-root node to non-root node) through \gls{rpl}'s \textit{Modes of Operation (MOP)}\cite{RFC6550}, which dictate how the downward routes are created. 

\gls{rpl} has five types of control messages; four of them have two versions (base and secure versions), and the last one has only a secure version. The secure version of \gls{rpl}'s control messages adds new unencrypted header fields and either a \gls{macc} or a digital signature field to the end of the base version, then encrypts the base part and the \gls{macc}\cite{RFC6550}.

\gls{dio} and \acrfull{dis} messages are used for the creation and maintenance of the \gls{dodag}\cite{RFC6550}. The root node starts the \gls{dodag} creation by multicasting a \gls{dio} message that contains the essential \gls{dodag} configurations and the root node's rank (the root node has the lowest rank in the \gls{dodag}). Upon receiving a \gls{dio} message, each node will select its \textit{preferred parent}, calculate its own rank, and multicast a new \gls{dio} with its calculated rank\cite{RFC6550, Raoof2018}. \acrshort{dis} messages are used to solicit \gls{dio} messages from node's neighbors when needed, e.g., a new node wants to join the networks or no \gls{dio} messages had arrived for a long time\cite{RFC6550}.

\gls{dao} and \glspl{daoack} messages are the backbones of the downward routes creation\cite{RFC6550}. The \gls{dao} contains path information about reachable nodes by its sender, and depending on \gls{rpl}'s \acrlong{mop2}, it will be used to create the downward routing table. Based on the \gls{dodag}'s configurations, a flag in the \gls{dao} message will mandate an acknowledgment (\gls{daoack} message) from the receiver.

\subsection{RPL's Security Mechanisms}
To secure the routing service, \gls{rpl} either relies on the security measures at the Link layer (i.e., IEEE 802.15.4\cite{STND802154}) or uses its own security mechanisms, resembled in three modes of security and an optional replay protection mechanism\cite{RFC6550, Perazzo2017}: 
The default mode for \gls{rpl} is the \textit{\textbf{Unsecured}} mode (\acrshort{umrpl}), where only the link-layer security is applied, if available. The second mode, the \textit{\textbf{Preinstalled}} secure mode (\gls{psmrpl}), which uses the preinstalled symmetrical encryption keys to secure \gls{rpl} control messages. Finally, the \textit{\textbf{Authenticated}} secure mode (\acrshort{asmrpl}) uses the preinstalled keys for the nodes to join the network, after which all routing-capable nodes have to acquire new keys from an authentication authority. To protect the routing service from replay attacks, \gls{rpl} uses Consistency Checks as an optional mechanism that can be used with either the preinstalled (\gls{psmrpl}rp) or authenticated mode (\acrshort{asmrpl}rp). In these checks, a special secure control message (\gls{cc} message) with non-repetitive nonce value is exchanged and used to assure no replay had occurred\cite{RFC6550}.

\section{Evaluation of RPL's Security Mechanisms under Attacks}\label{RPLEval}
In this paper, \gls{rpl} performance is evaluated against four attacks\cite{Raoof2018, Wallgren2013}: the Blackhole, the Selective-Forward, the Neighbor, and the Wormhole attacks. Experiments were conducted with \gls{rpl} in both \acrshort{umrpl} (vanilla ContikiRPL) and \gls{psmrpl} (as in Perazzo \textit{et al.}\cite{Perazzo2017} implementation). For the latter, we evaluated \gls{rpl} with and without the optional replay protection mechanism.

\subsection{Evaluation Setup}\label{EvalSetup}
\begin{figure}[!t]
\centering
\includegraphics[height=70mm, width=75mm]{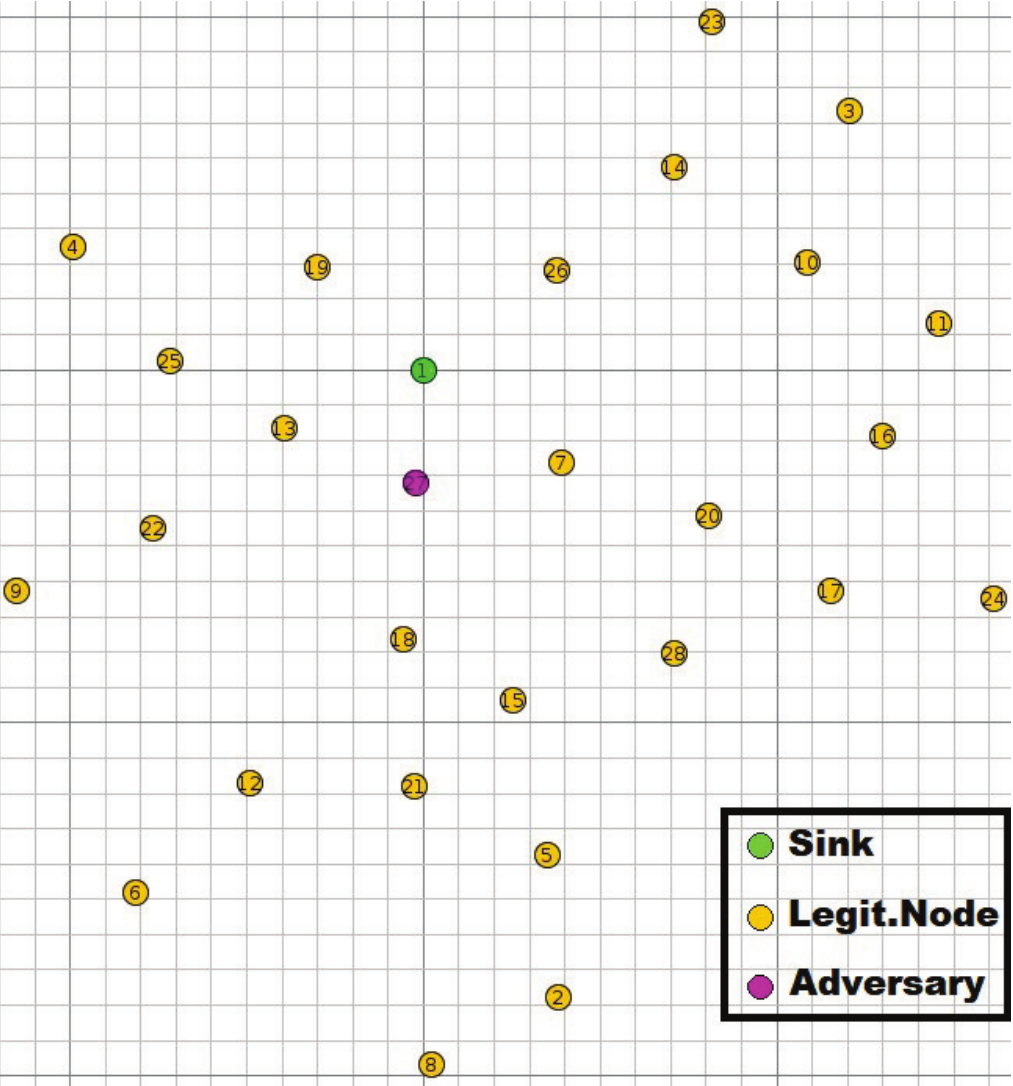}
\caption{Network topology for the Blackhole, Selective-Forward, and Neighbor attacks scenarios (better viewed in colors.)}
\label{fig_1}
\end{figure}

\begin{figure}[!t]
	\centering
	\includegraphics[height=70mm, width=75mm]{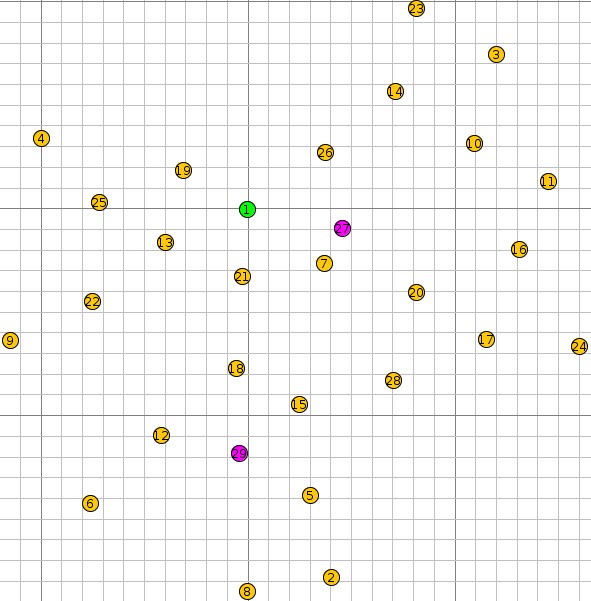}
	\caption{Network topology for the Wormhole attack scenario.}
	\label{fig_10}
\end{figure}

Cooja, the simulator for Contiki OS\cite{ContikiOSRef}, was used for all the simulations (with simulated motes). Fig.\ref{fig_1} shows the topology used in our evaluation for the Blackhole, Selective-Forward, and Neighbor attacks, while Fig.\ref{fig_10} shows the one used to evaluate the Wormhole attack (as two adversaries are needed). A list of simulation parameters is provided in Table \ref{table_1}.


\begin{table}[!t]
	\caption{List of Simulation Parameters}
	\label{table_1}
	\begin{tabular}{ll}
		\toprule
		\textbf{Description}&\textbf{Value}\\
		\midrule
		No. of sim. sets&
		\begin{minipage}{0.38\columnwidth}
			Two: one for each \gls{rdc} protocol (See \S\ref{ImptCh})
		\end{minipage} \\
		\midrule
		No. of experiments per set&
		\begin{minipage}{0.38\columnwidth}
			Four (See \S\ref{AttModANDSen})
		\end{minipage} \\
		\midrule
		No. of scenarios per experiment&4 (ContikiMAC) / 5 (NullRDC)\\
		\midrule
		No. of sim. rounds per scenario / time&10 rounds / 20 min. per round\\
		\midrule
		Node Positioning&Random (three clusters)\\
		\midrule
		Deployment area&290m W x 310m L\\
		\midrule
		Number of nodes (adversary included)&
		\begin{minipage}{0.38\columnwidth}
		28 (ContikiMAC) / 29 (NullRDC)
		\end{minipage} \\
		\midrule
		Sensor nodes type&
		\begin{minipage}{0.38\columnwidth}
			Arago Sys. Wismote mote
		\end{minipage}\\
		\midrule
		DATA transmission rate& 
		\begin{minipage}{0.38\columnwidth}
			$\simeq$ 1 packet per minute per legitimate node
		\end{minipage}\\
		\bottomrule
	\end{tabular}
\end{table}

Both topologies represent a single \gls{dodag} network with one root or sink node (the green node). The minimum number of adversaries required for each attack was used to reduce the complexity of the observed metrics. For the Blackhole, Selective-Forward, and Neighbor attacks, node (27) was used as an adversary and positioned near the sink node, which would introduce the most prominent effect of the three attacks\cite{Pu2018a,Wallgren2013,Mayzaud2016}. For the Wormhole attack, two adversaries (nodes 27 and 29) were used and positioned to create a wormhole between the node cluster (1, 7, 20, and 26) and the targeted nodes. The targeted nodes for all the attacks are (2, 5, 6, 8, 12, 15, 18, 21, and 28), with node (28) providing an alternative path for the targeted nodes to send their packets toward the sink. Having an alternative path is crucial to our experiments to examine how the self-healing mechanisms of \gls{rpl} will respond to the attacks.

Note that we tried to implement the simulations using Zolertia Z1 motes\cite{Z1} (each has 8KB RAM and 92KB Flash memory) to compare our results to that of \cite{Raoof2019}. However, enabling the replay protection mechanism of \gls{rpl} in our simulation caused the mote to always run out of RAM, rendering the simulation impractical. Hence, we moved to the more powerful Wismote motes (each has 16KB RAM and a 256KB Flash memory\cite{Wismote}).

\subsection{Assumptions}\label{EvalAssump}
The following assumptions were used in our evaluation: \gls{rpl} uses the default \gls{of}, namely the \gls{mrhof}\cite{RFC6719}. To keep the focus on \gls{rpl} at the Network layer, we assumed neither security measures nor encryption was enabled at the Link layer. All the attacks were implemented at the Network layer.

Our data traffic model is a deterministic one that mimics a typical sensing-\gls{iot} network, where nodes send their sensor readings toward the root node at predetermined periods. In our model, only the legitimate nodes send data packets toward the root, each sending one packet per minute, while the adversaries only participate in the \gls{dodag} formation without sending any data packets.

The results obtained from the simulations were averaged over ten rounds for each scenario with a 95\% confidence level.

\subsection{Adversary Model and Attack Scenarios}\label{AttModANDSen}
For each \gls{rdc} protocol (see \S\ref{ImptCh}), we conducted a \textit{set} of four experiments: the first three experiments (\gls{rpl} in \acrshort{umrpl}, \gls{rpl} in \gls{psmrpl}, and \gls{rpl} in \gls{psmrpl}rp) have an \textit{internal} adversary, who participates in the creation of the topology from the beginning (and has the preinstalled encryption keys in the 2$^{nd}$ and 3$^{rd}$ experiments). The fourth experiment (\gls{rpl} in \gls{psmrpl}) uses an \textit{external} adversary who runs \gls{rpl} in \acrshort{umrpl} and does not have the knowledge of the secure versions of \gls{rpl}'s control messages, while the legitimate nodes run \gls{rpl} in \gls{psmrpl}. Table \ref{Exp_Sum} lists the settings for these experiments.

For the attacks themselves, we have five scenarios:
\begin{enumerate}
	\item \textit{No Attack}: the adversary works as a fully legitimate node.
	\item \textit{Blackhole Attack (BH)}: the adversary drops all types of traffic coming through, including \gls{rpl} control messages and data packets \cite{Raoof2018}. In our evaluation, the adversary will keep its radio operational according to the \gls{rdc} protocol, but it will simply discard any incoming or outgoing frames.
	\item \textit{Selective-Forward Attack (SF)}: the adversary drops any non-\gls{rpl} packets, including data. However, only \gls{rpl} control messages will be processed as normal and passed \cite{Pu2018a}. Similarly to BH, our simulation of the adversary will keep its radio operational, but it will check the "\textit{Type}" field of any incoming or outgoing \gls{icmp} packet to see if it holds an \gls{rpl} control message (\gls{rpl}'s \textit{Type} is 155). If the packet is an \gls{rpl} control message, it will be processed as usual and passed. All other types of packets will be discarded.
	\item \textit{Neighbor Attack (NA)}: the adversary will pass any \gls{dio} message it receives from its neighbors without any processing or modification \cite{Le2013a}. This will create the illusion of having the original sender in the range of the victim nodes. Our simulation of this adversary is a simple one: while operating as a legitimate node, whenever the adversary receives a \gls{dio} message (even if it was not addressed to it), it will multicast an exact copy of the received message to its sub-\gls{dodag} before processing the message as usual.
	\item Out-of-Band \textit{Wormhole Attack (WH)}: two adversaries use an out-of-band link to forward \gls{rpl} control messages from legitimate nodes between the two locations where the adversaries reside\cite{Pongle2015a, Raoof2018}. This scenario is available only in the NullRDC set of experiments, see \S\ref{ImptCh}.
\end{enumerate}

In the BH, SF, and NA scenarios, the adversary always starts as a legitimate node, tries to join the network, and actively participates in the creation and maintenance of the \gls{dodag}. Then, it works as a legitimate node for two minutes (to assure full integration with the network) before launching the attack afterward. For the Wormhole attack, the two adversaries are always in promiscuous mode and never participate in the \gls{dodag}.

The choice of these attacks was based on the fact that they have a minimum cost for the adversary to launch them, as they require little or no processing of \gls{rpl}’s messages. At the same time, the effect of these attacks can be significant on the network.

It is worth mentioning that our simulation of the Wormhole attack is based upon the work in \cite{Perazzo2018}. The authors implemented an out-of-band wormhole on a real testbed, with a wired link between the adversaries. Each adversary operates in the \textit{promiscuous mode}, sniffs all types of frames, sends the sniffed frames through the wired link, and replays the frames it received from the wired link. However, our implementation differs from theirs in a few points:
\begin{enumerate}
	\item Our implementation is simulation-based and is conducted in Cooja. We use the host computer to emulate a fast link between the adversaries.
	\item The wormhole is implemented at the Network layer level in order to detect and replay \gls{rpl}'s control messages only. In addition, the adversaries can identify the secure versions of \gls{rpl}'s control messages.
	\item The adversaries use a multi-buffer approach for the packets received from the radio and for the packets awaiting the replay. This approach accelerates the operation of the adversaries when there are many neighbors, and makes sure that all forwarded packets are replayed without dropping any of them.
\end{enumerate}
\begin{table}[!t]
	\caption{Experiments summary}
	\label{Exp_Sum}
	\centering
	\begin{tabular}{c|c|c|c}
		\toprule
		\textbf{Experiment} & \textbf{Secure Mode} & \textbf{Replay Protection} & \textbf{Adversary Type}\\
		\toprule
		UM-I & $\times$ & $\times$ & Internal (I)\\
		\midrule
		PSM-I & \checkmark & $\times$ & Internal (I)\\
		\midrule
		PSMrp-I & \checkmark & \checkmark & Internal (I)\\
		\midrule
		PSM-E & \checkmark & $\times$ & External (E)\\
		\bottomrule
	\end{tabular}
\end{table}

\subsection{Implementation Challenges}\label{ImptCh}
Contiki OS \cite{ContikiOSRef} divides the Link layer into three sub-layers: the \textit{\gls{macl}} sub-layer, which is responsible for addressing, sequencing, and retransmissions; the \textit{FRAMER} sub-layer that is responsible for creating and parsing of frames; and the \textit{\gls{rdc}} sub-layer that controls the radio component. Currently, Contiki \gls{os} comes with several \gls{rdc} protocols, with the most used ones are the ContikiMAC\cite{ContikiMAC2011} and NullRDC.

ContikiMAC is the default setting for RDC protocol in Contiki OS. Here, the radio is kept off most of the time, with the protocol waking up the radio periodically to check for transmissions. If a transmission is detected, the radio will be kept on long enough to receive the frame, send an ACK to the sender (if it was accepted)\cite{Barnawi2019, ContikiMAC2011}, then the radio is turned off. Similarly, the sender will turn on the radio, probe the channel, perform several attempts to transmit a frame, and wait for either an ACK (which dictates a successful transmission) or reach a threshold that means a failed transmission\cite{ContikiMAC2011}. Either way, the radio is turned off afterward. ContikiMAC protocol is proved to be very efficient with power consumption, at the expense of having longer \gls{e2e} latency\cite{Barnawi2019}.

On the other hand, the NullRDC protocol keeps the radio always on and does not perform frequent channel probing, which means lower \gls{e2e} latency and a smaller number of retransmissions at the expense of higher power consumption\cite{Barnawi2019}.

During our implementation of the Wormhole attack using the ContikiMAC protocol, we found that the messages forwarded through the wormhole were replayed very late by the adversaries; hence, those messages got ignored by the legitimate nodes. A further investigation showed that a mix of simulation environment latency and the lengthy sending procedure of ContikiMAC are the culprits for such late replay. Several trials were made to reduce simulation latency (e.g., reducing output text, using faster host, etc.) and accelerate ContikiMAC sending procedure; all have failed.

However, implementing the Wormhole attack using the NullRDC protocol proved to be working perfectly. Since the sending procedure is much simpler than that of ContikiMAC, the Wormhole attack performed as expected, without any added latency and resulting in full disruption to the routing topology (as explained later). Since the power consumption, in this case, is dominated by the high usage of the always-on radio (almost fixed at 122 milliwatts), we are not able to evaluate the effect of the investigated attacks on power consumption using the NullRDC protocol.

For that reason, only the NullRDC experiment set evaluates the Wormhole attack, omitting the power consumption metric.
\begin{figure*}[ht]
	\centering
	\subfloat[Average packet delivery rate (\gls{pdr}).]{\includegraphics[height=4.30cm, width=.2166\linewidth]{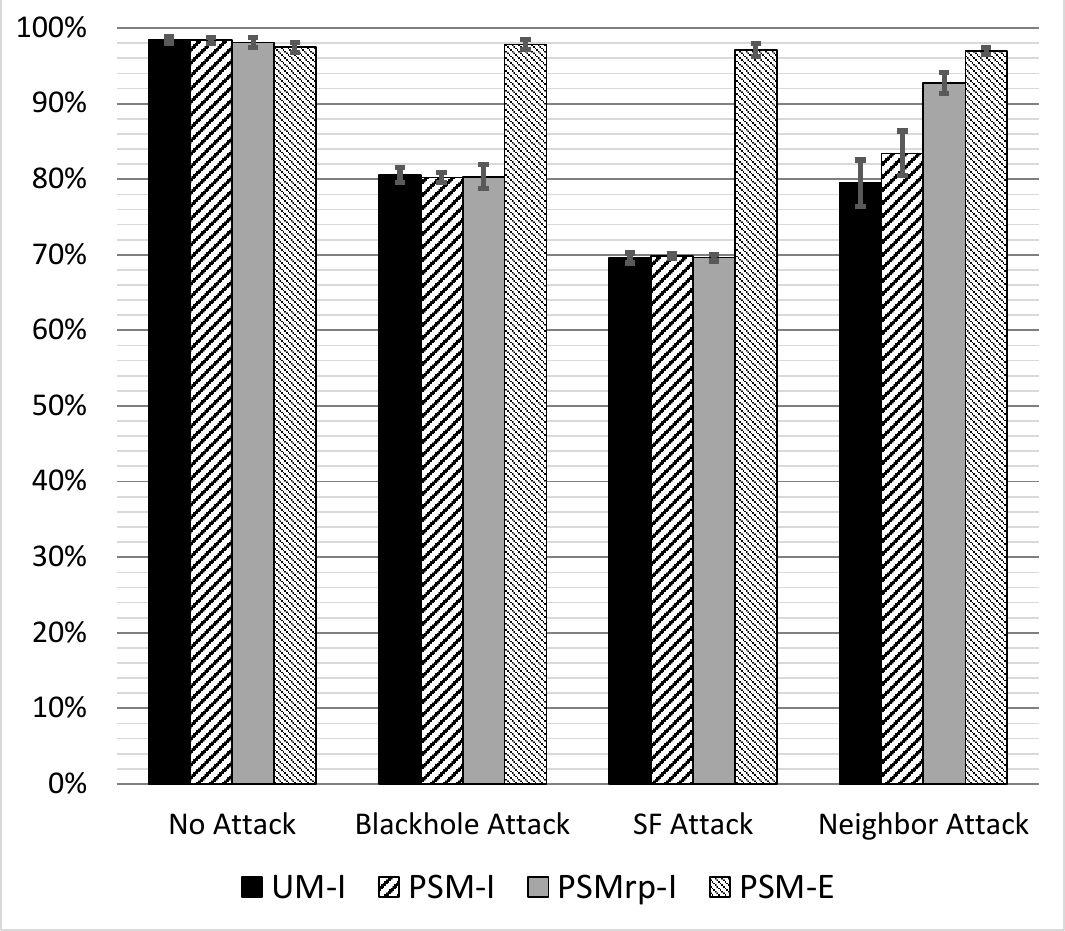}
		\label{fig_2a}}
	\hfil
	\subfloat[Average network \gls{e2e} latency.]{\includegraphics[height=4.30cm, width=.2166\linewidth]{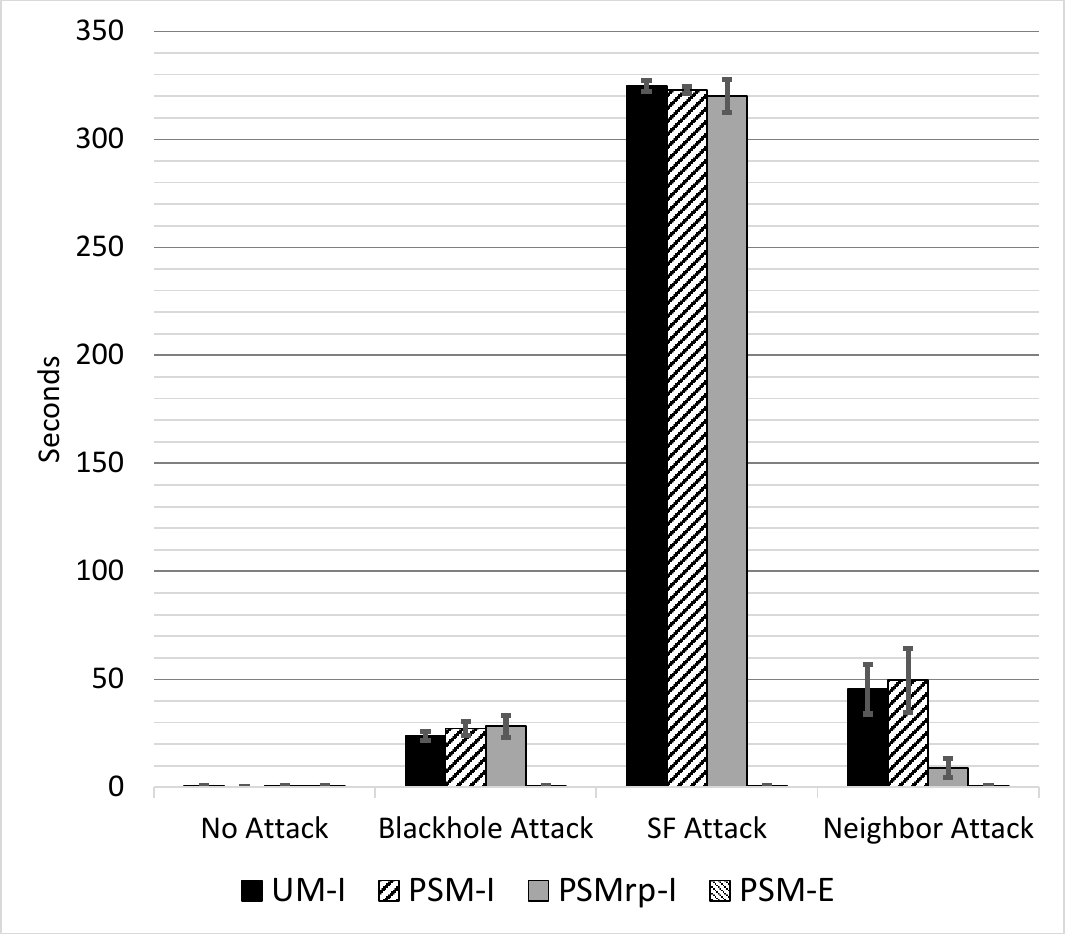}
		\label{fig_2b}}
	\hfil
	\subfloat[Average network power consumption, per received packet.]{\includegraphics[height=4.30cm, width=.2166\linewidth]{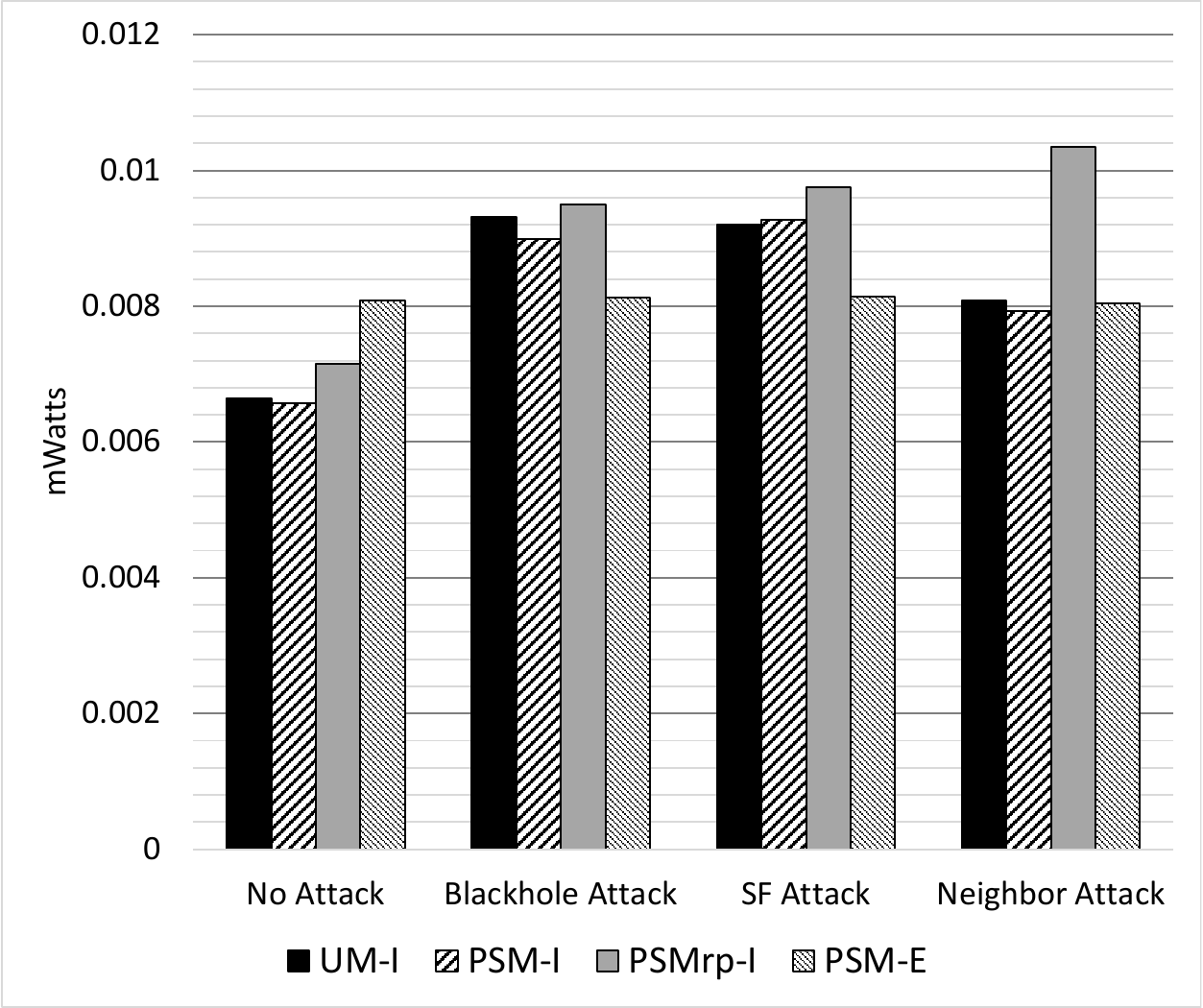}
		\label{fig_2d}}
	\hfil
	\subfloat[Exchanged \gls{rpl} control messages, per legitimate node.]{\includegraphics[height=4.3cm, width=.35\linewidth]{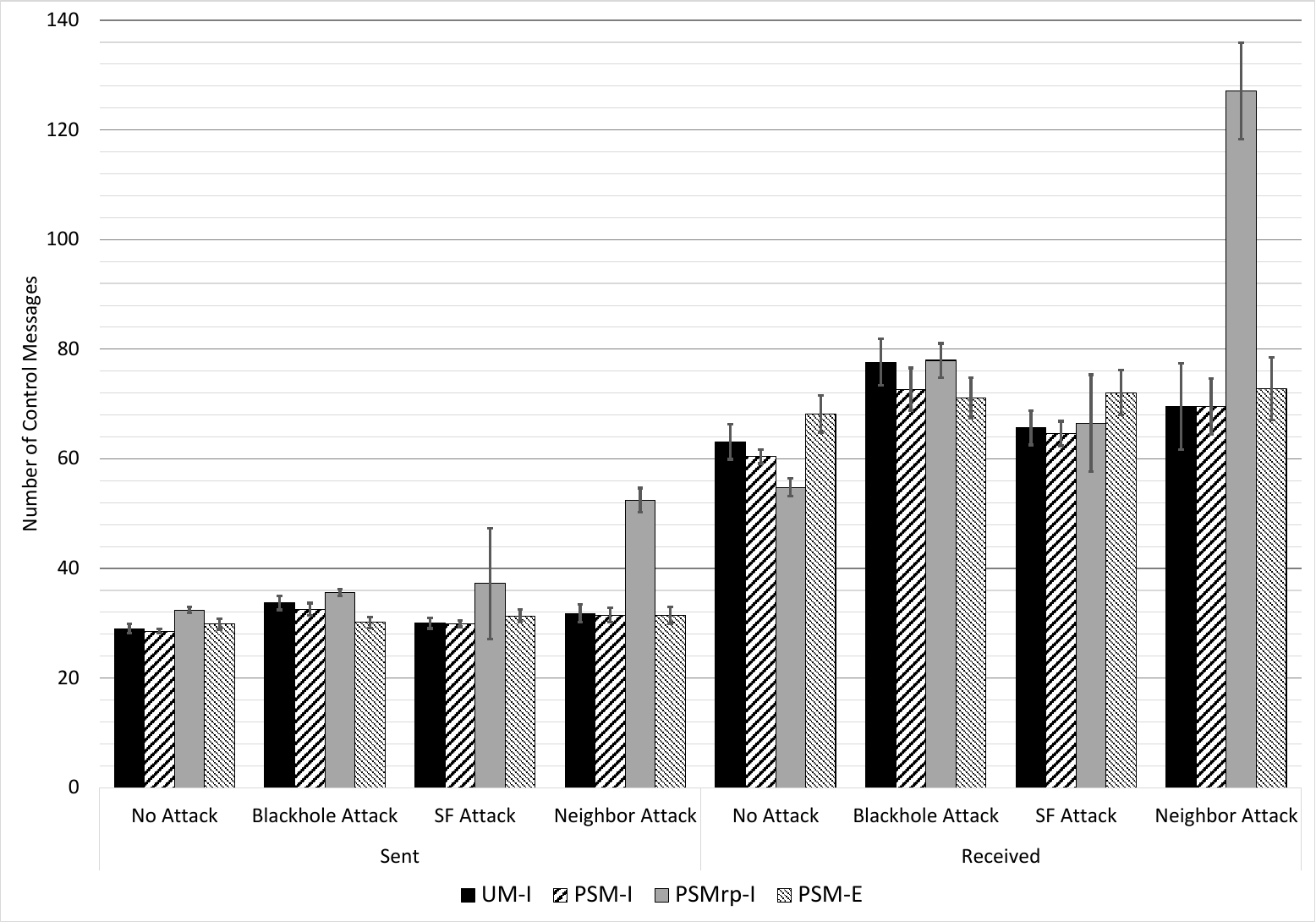}
		\label{fig_2c}}
	\hfil
	\caption{Simulation results for the four experiments (three attacks scenarios), using ContikiMAC RDC protocol. (UM: unsecured mode, PSM: preinstalled secure mode, PSMrp: preinstalled secure mode with replay protection, I: internal adversary, E: external adversary.)}
	\label{fig_2}
\end{figure*}
\begin{figure*}[h]
	\centering
	\subfloat[Average packet delivery rate (\gls{pdr}).]{\includegraphics[height=4.30cm, width=.2166\linewidth]{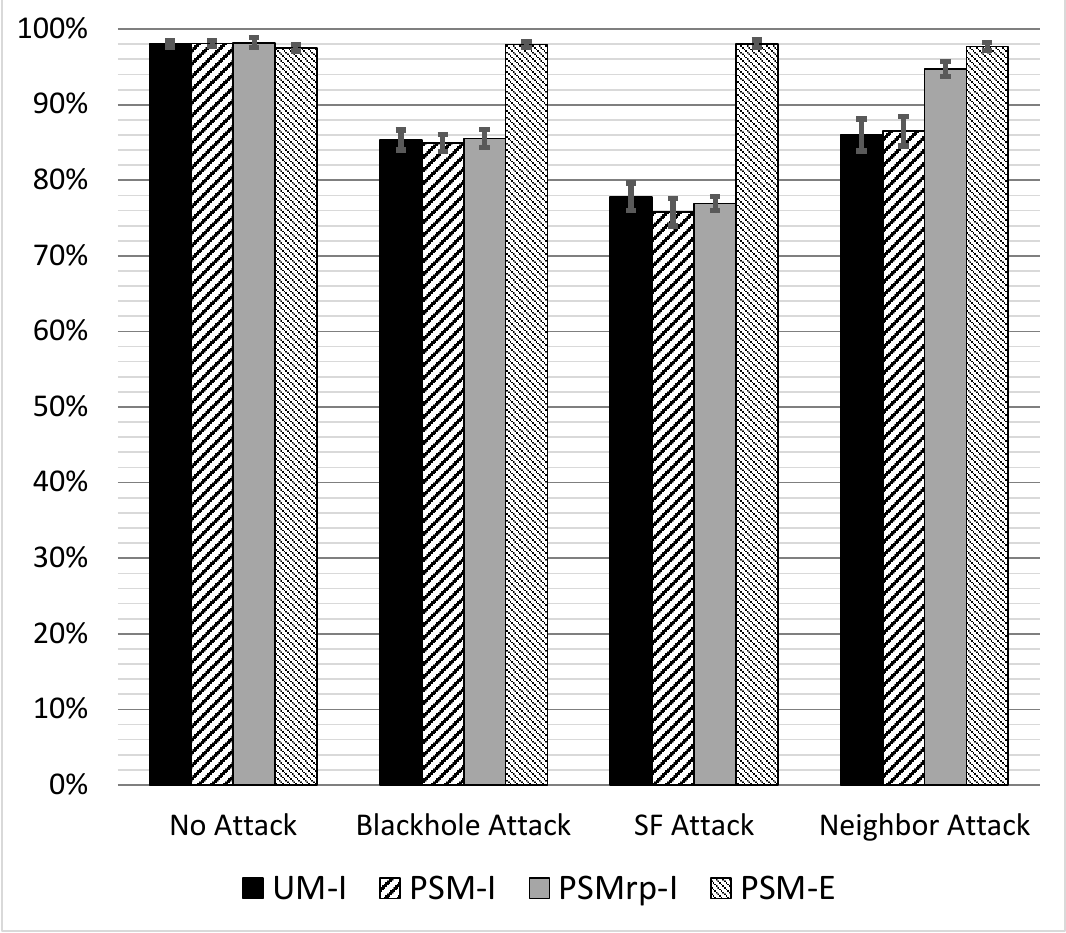}%
		\label{fig_5a}}
	\hfil
	\subfloat[Average network \gls{e2e} latency.]{\includegraphics[height=4.30cm, width=.2166\linewidth]{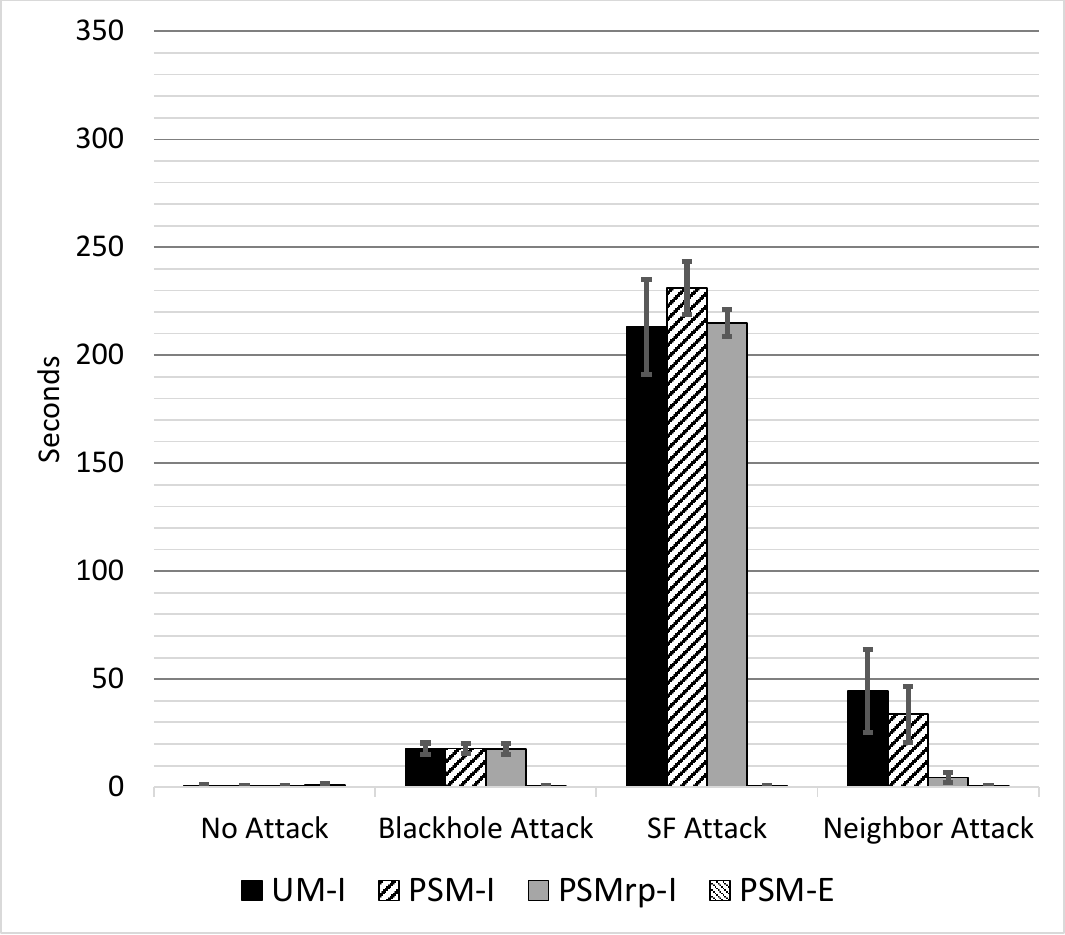}%
		\label{fig_5b}}
	\hfil
	\subfloat[Average network power consumption, per received packet.]{\includegraphics[height=4.30cm, width=.2166\linewidth]{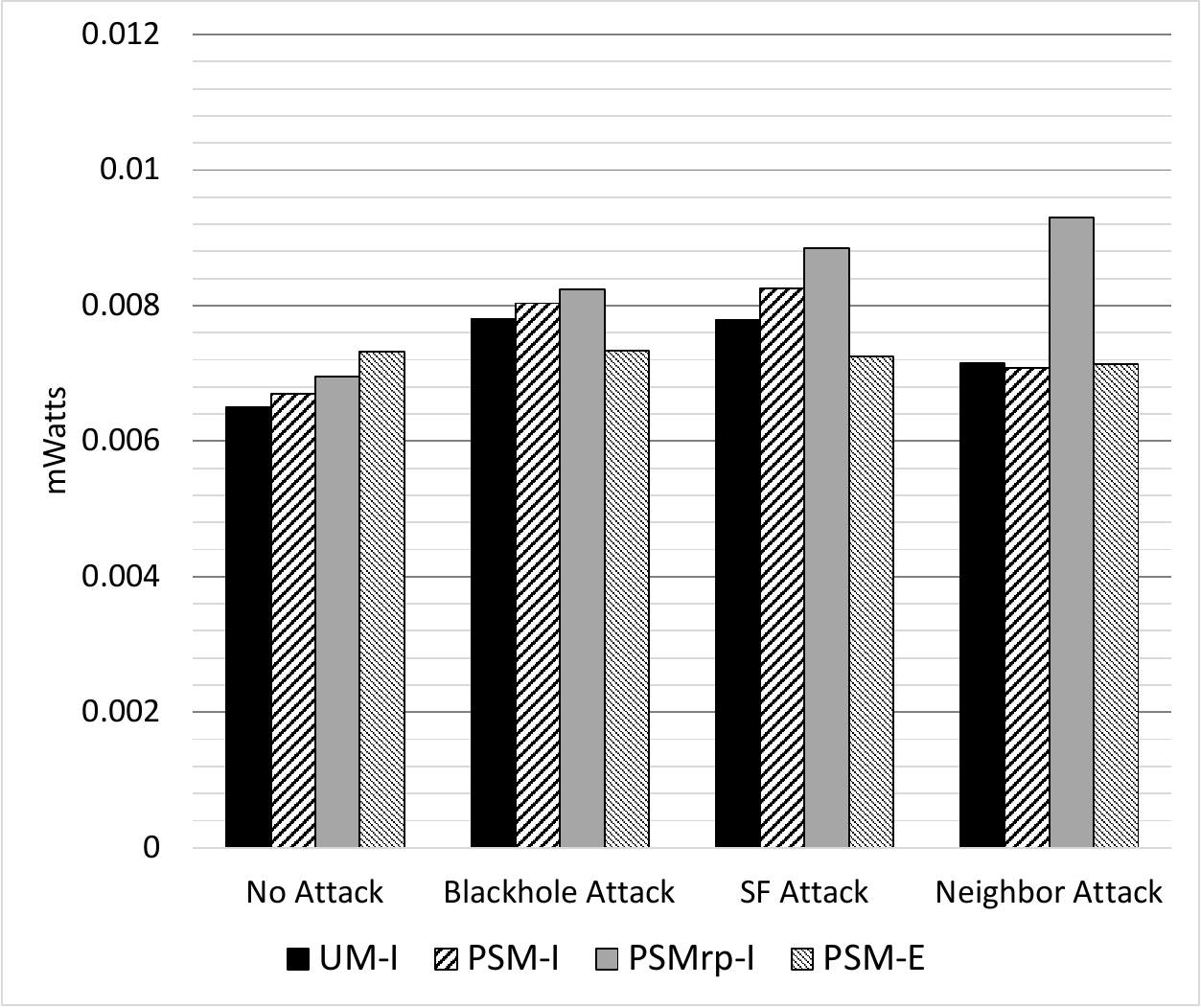}%
		\label{fig_5c}}
	\hfil
	\subfloat[Exchanged \gls{rpl} control messages, per legitimate node.]{\includegraphics[height=4.3cm, width=.35\linewidth]{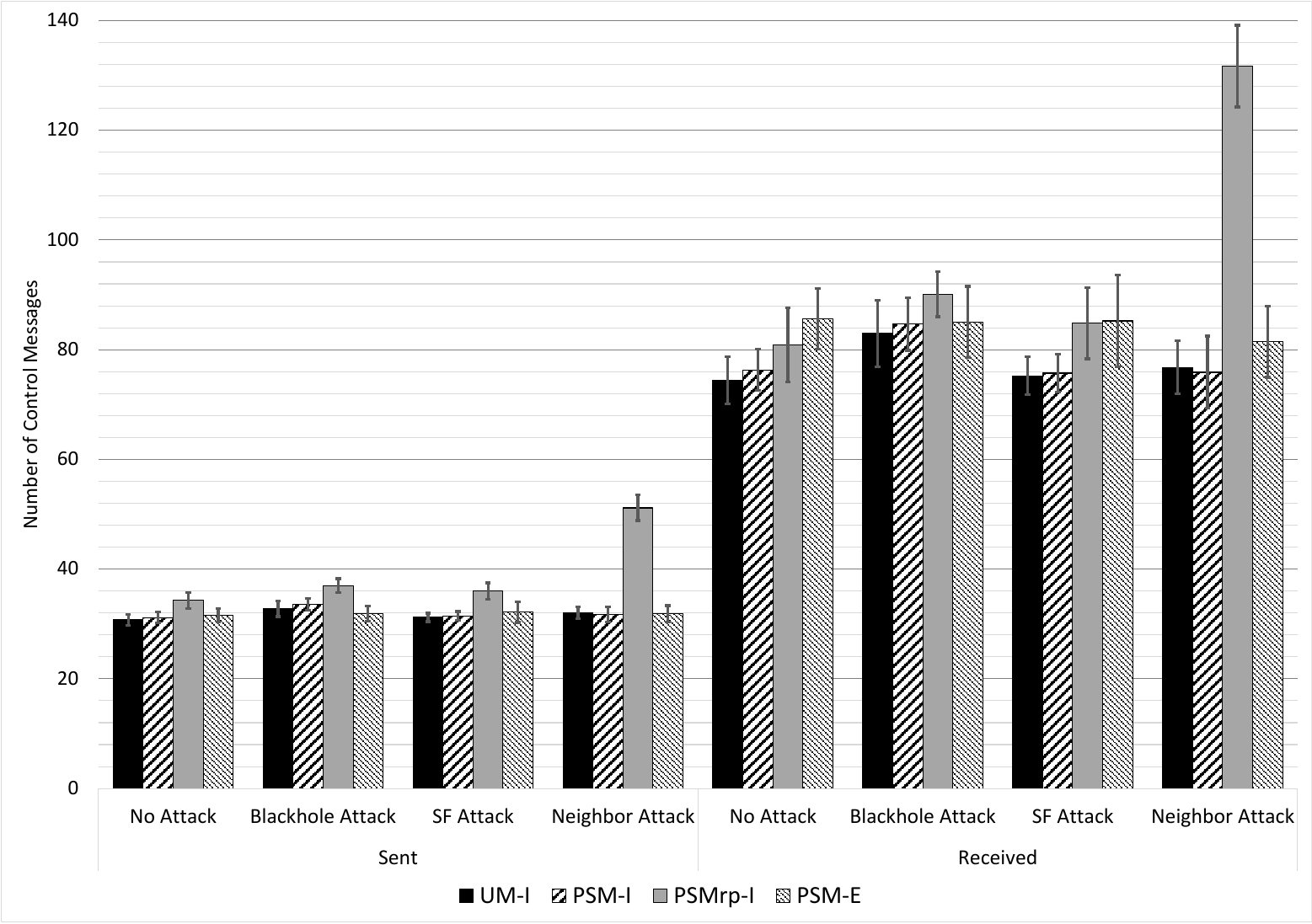}%
		\label{fig_5d}}
	\hfil
	\caption{Simulation results for the first suggestion (having more routes toward the root node), using ContikiMAC RDC protocol.}%
	\label{fig_5}
\end{figure*}
\begin{figure*}[!h]
	\centering
	\subfloat[Average packet delivery rate (\gls{pdr}).]{\includegraphics[height=4.30cm, width=.2166\linewidth]{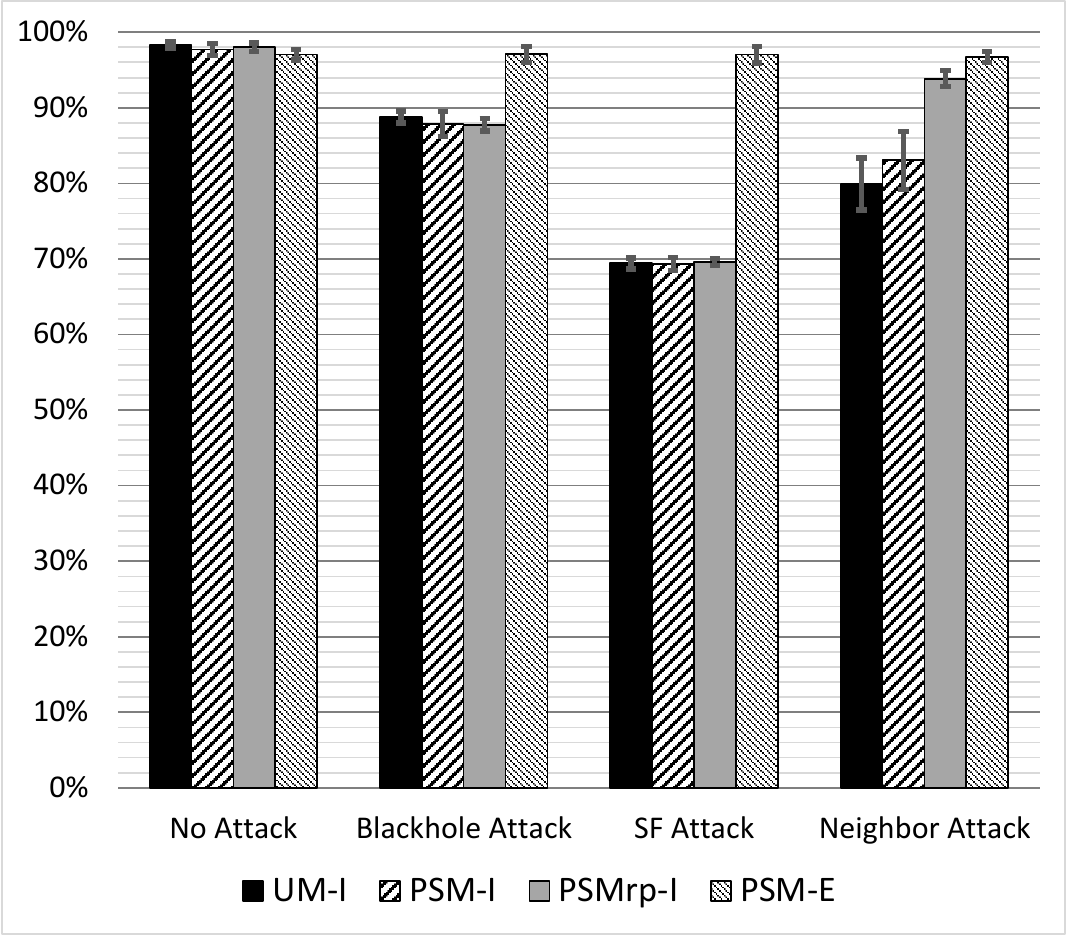}%
		\label{fig_6a}}
	\hfil
	\subfloat[Average network \gls{e2e} latency.]{\includegraphics[height=4.30cm, width=.2166\linewidth]{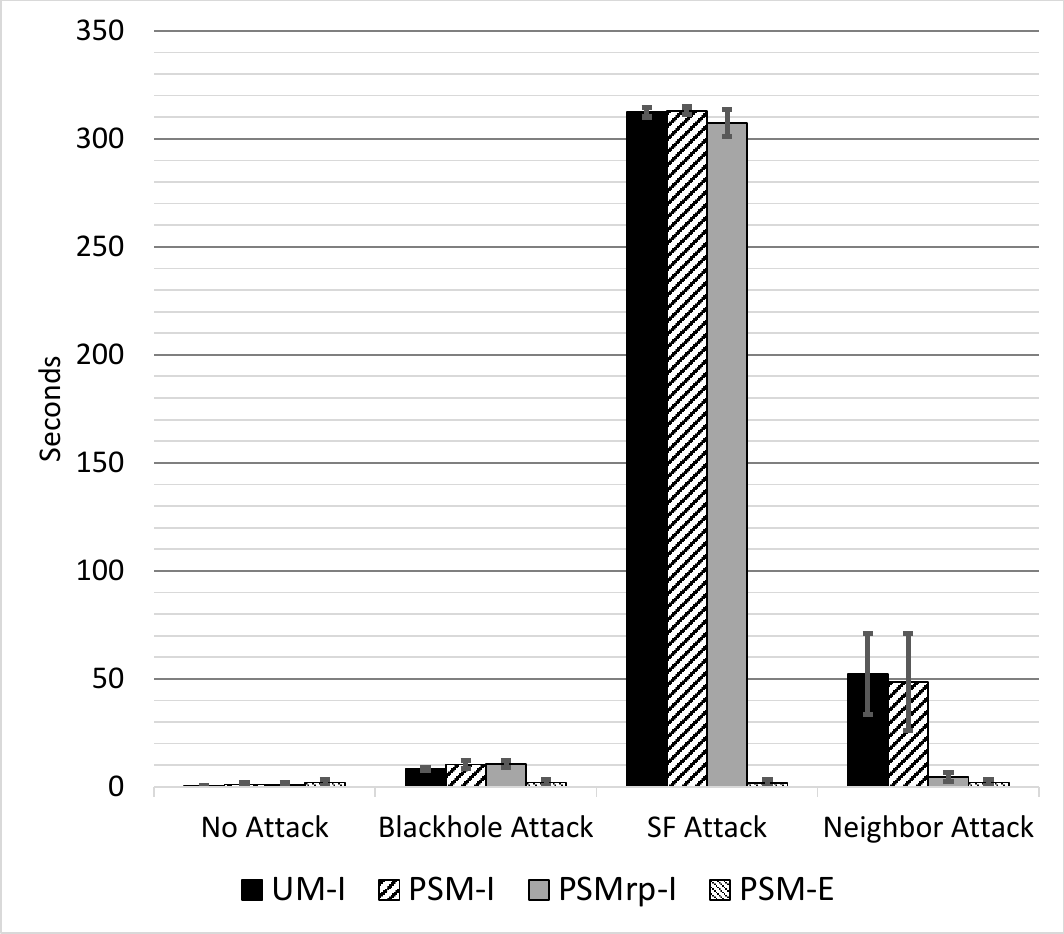}%
		\label{fig_6b}}
	\hfil
	\subfloat[Average network power consumption, per received packet.]{\includegraphics[height=4.30cm, width=.2166\linewidth]{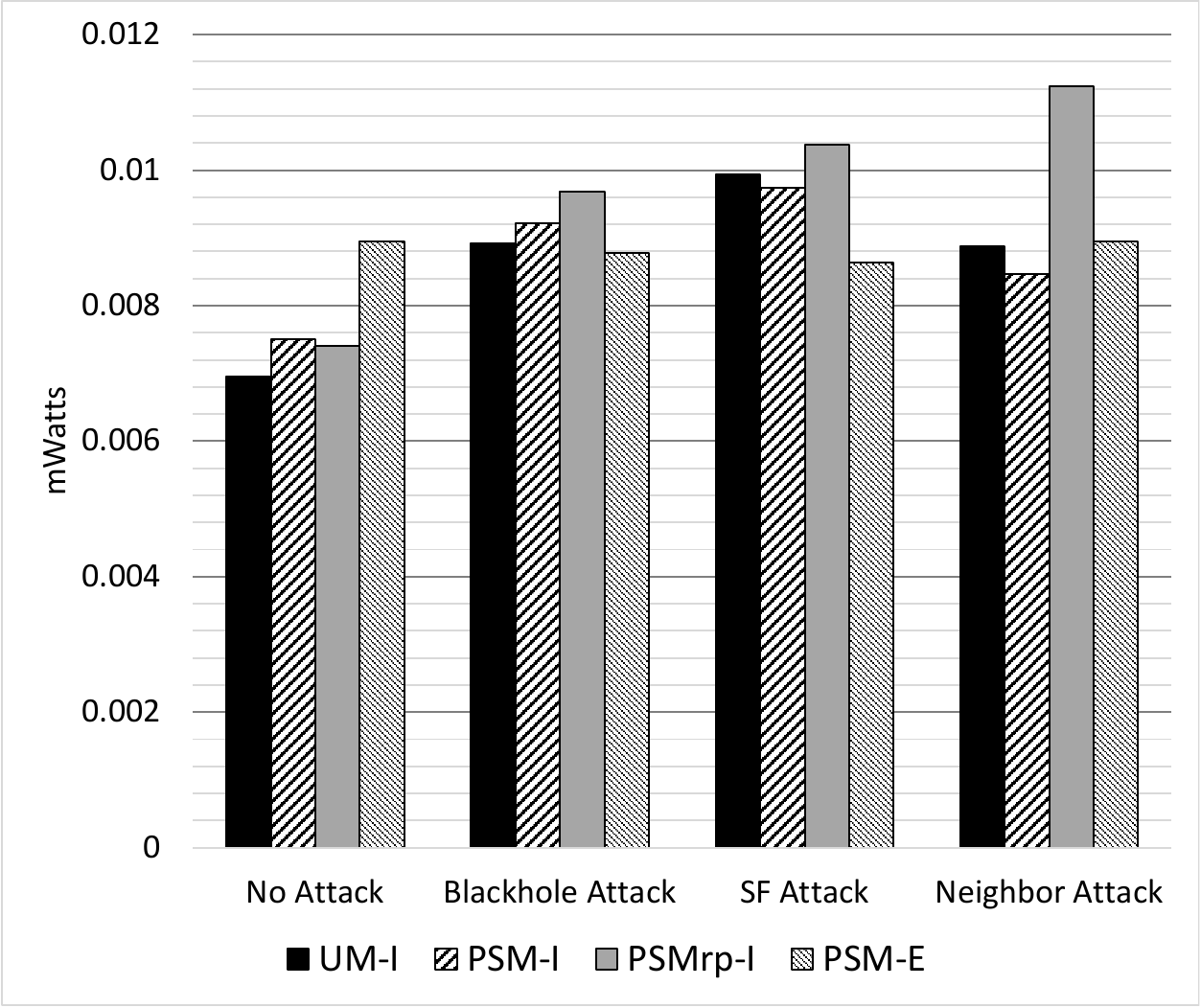}%
		\label{fig_6c}}
	\hfil
	\subfloat[Exchanged \gls{rpl} control messages, per legitimate node.]{\includegraphics[height=4.3cm, width=.35\linewidth]{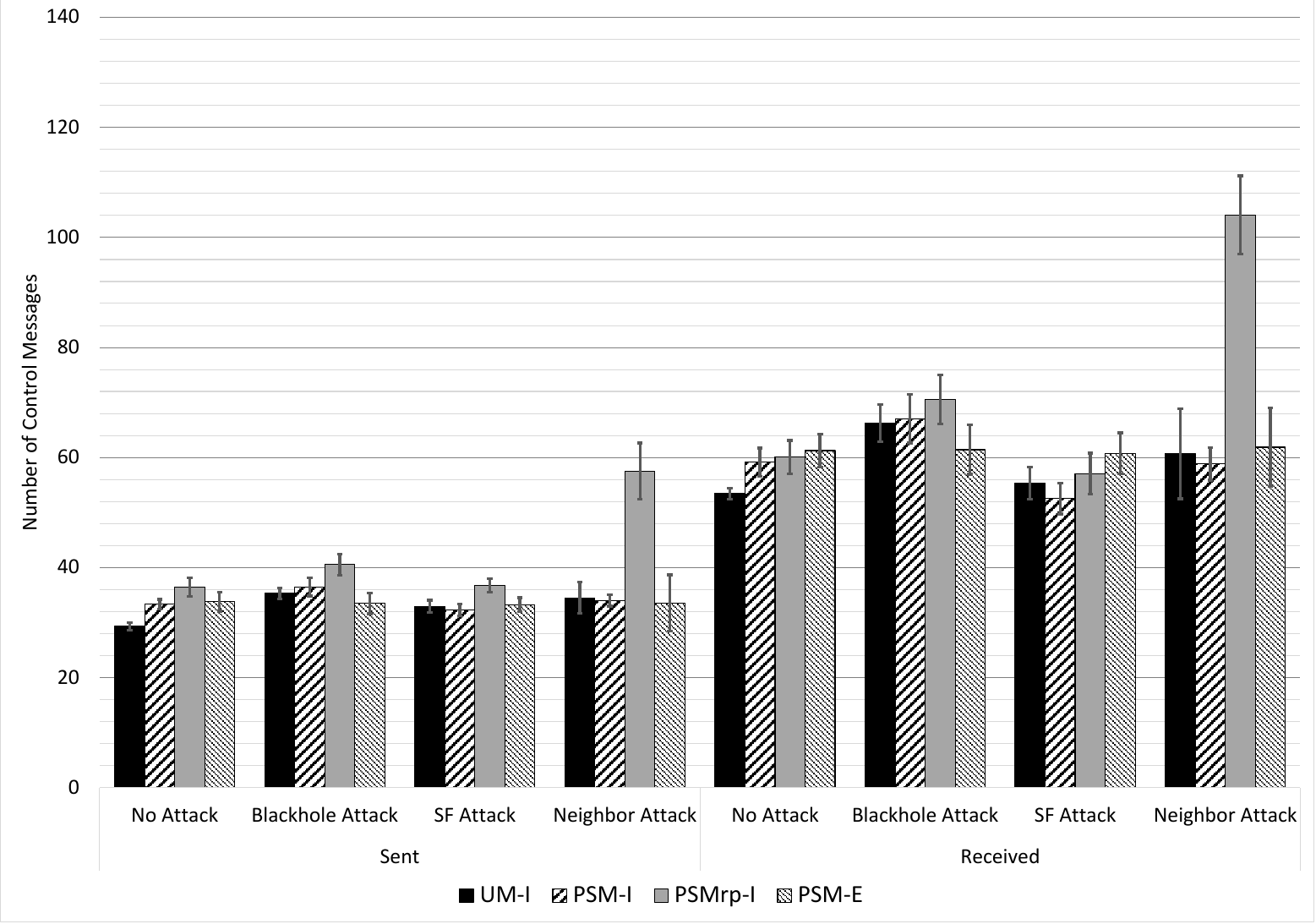}%
		\label{fig_6d}}
	\hfil
	\caption{Simulation results for the second suggestion (reducing the timeout value for declaring a parent as dead), using ContikiMAC RDC protocol.}%
	\label{fig_6}
\end{figure*}

\begin{figure*}[th]
	\centering
	\subfloat[Average packet delivery rate (\gls{pdr}).]{\includegraphics[scale=.29]{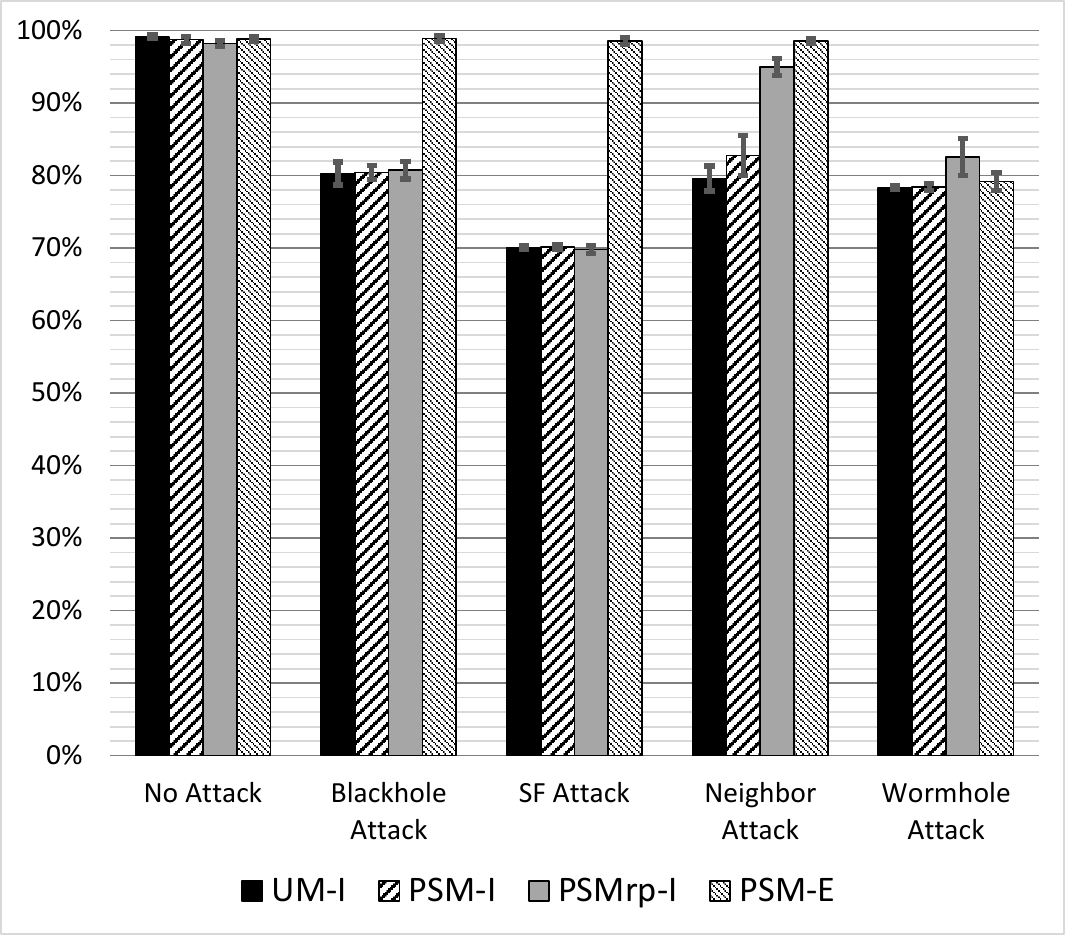}%
		\label{fig_7a}}
	\hfil
	\subfloat[Average network \gls{e2e} latency.]{\includegraphics[scale=.29]{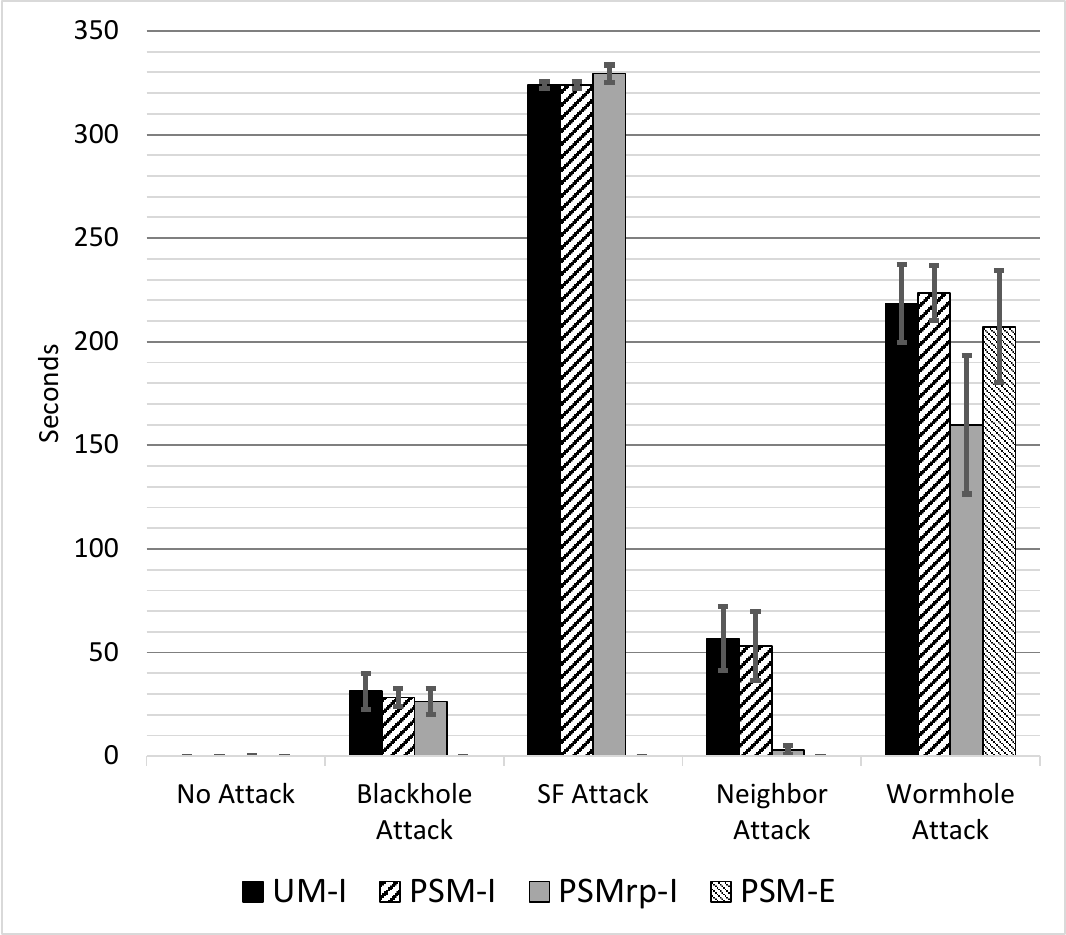}%
		\label{fig_7b}}
	\hfil
	\subfloat[Exchanged \gls{rpl} control messages, per legitimate node.]{\includegraphics[scale=.25]{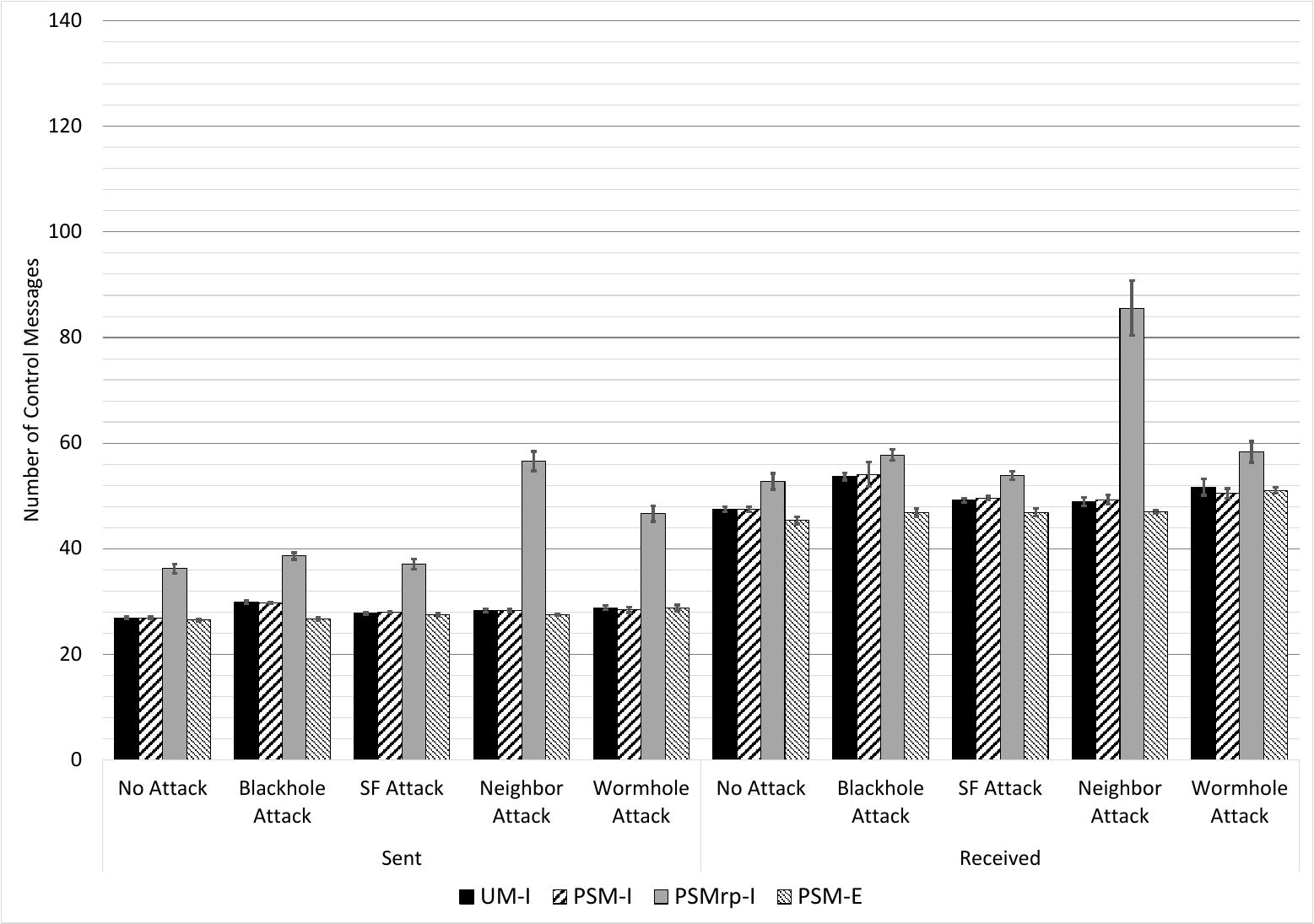}%
		\label{fig_7c}}
	\hfil
	\caption{Simulation results for the four experiments (four attacks scenarios), using NullRDC RDC protocol.}
	\label{fig_7}
\end{figure*}
\begin{figure*}[h]
	\centering
	\subfloat[Average packet delivery rate (\gls{pdr}).]{\includegraphics[scale=.29]{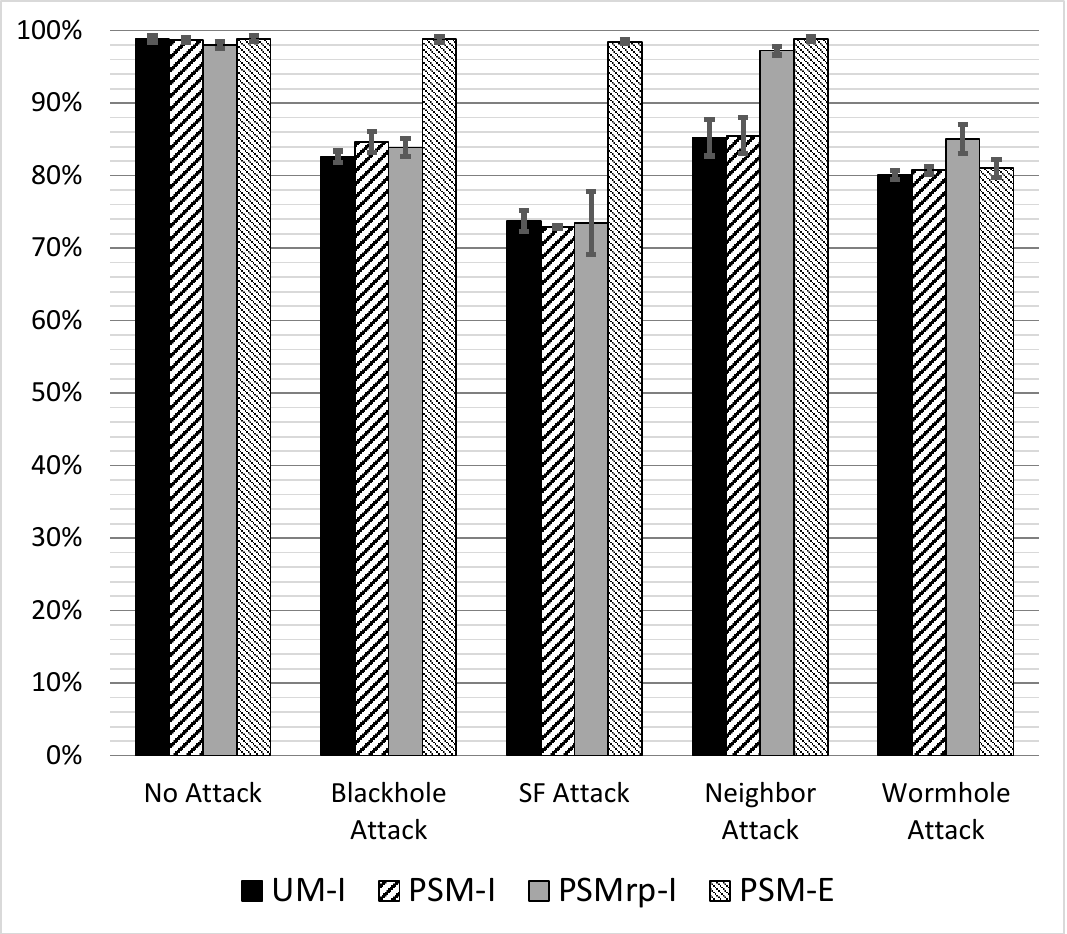}%
		\label{fig_8a}}
	\hfil
	\subfloat[Average network \gls{e2e} latency.]{\includegraphics[scale=.29]{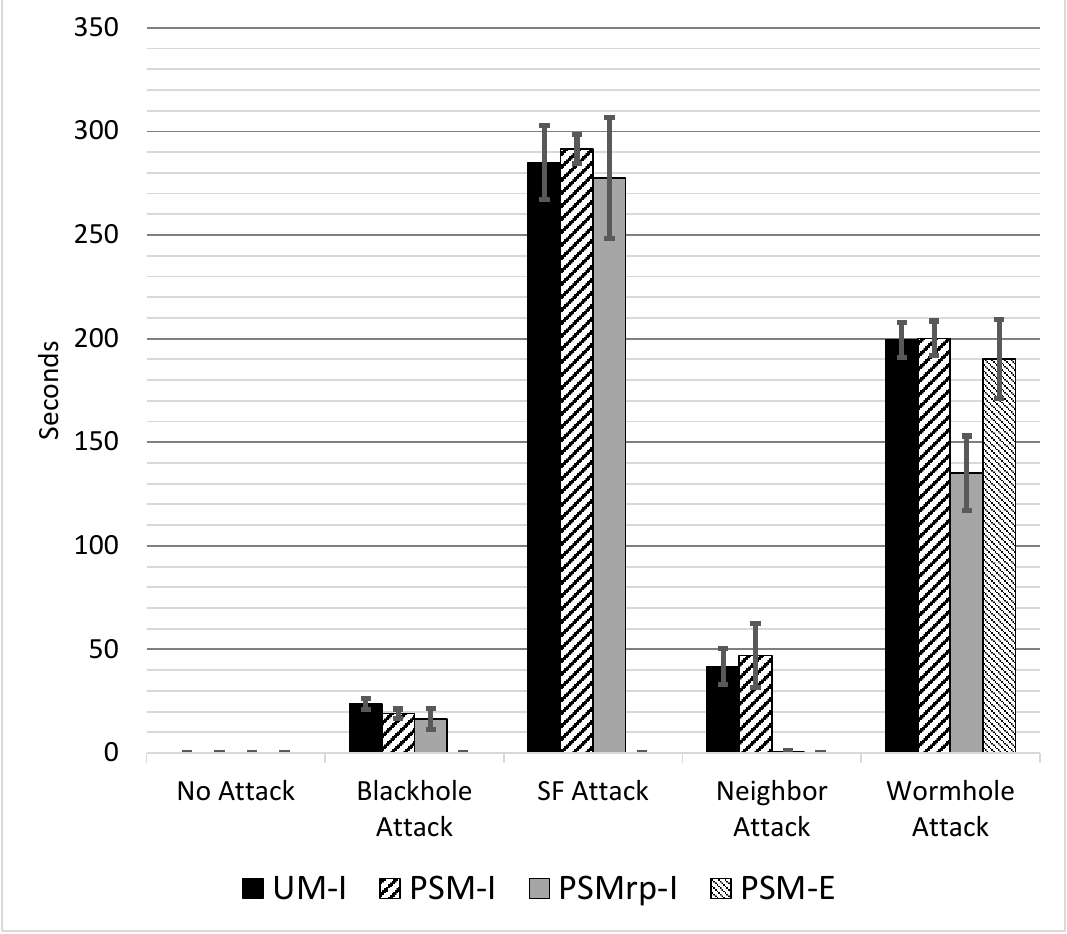}%
		\label{fig_8b}}
	\hfil
	\subfloat[Exchanged \gls{rpl} control messages, per legitimate node.]{\includegraphics[scale=.25]{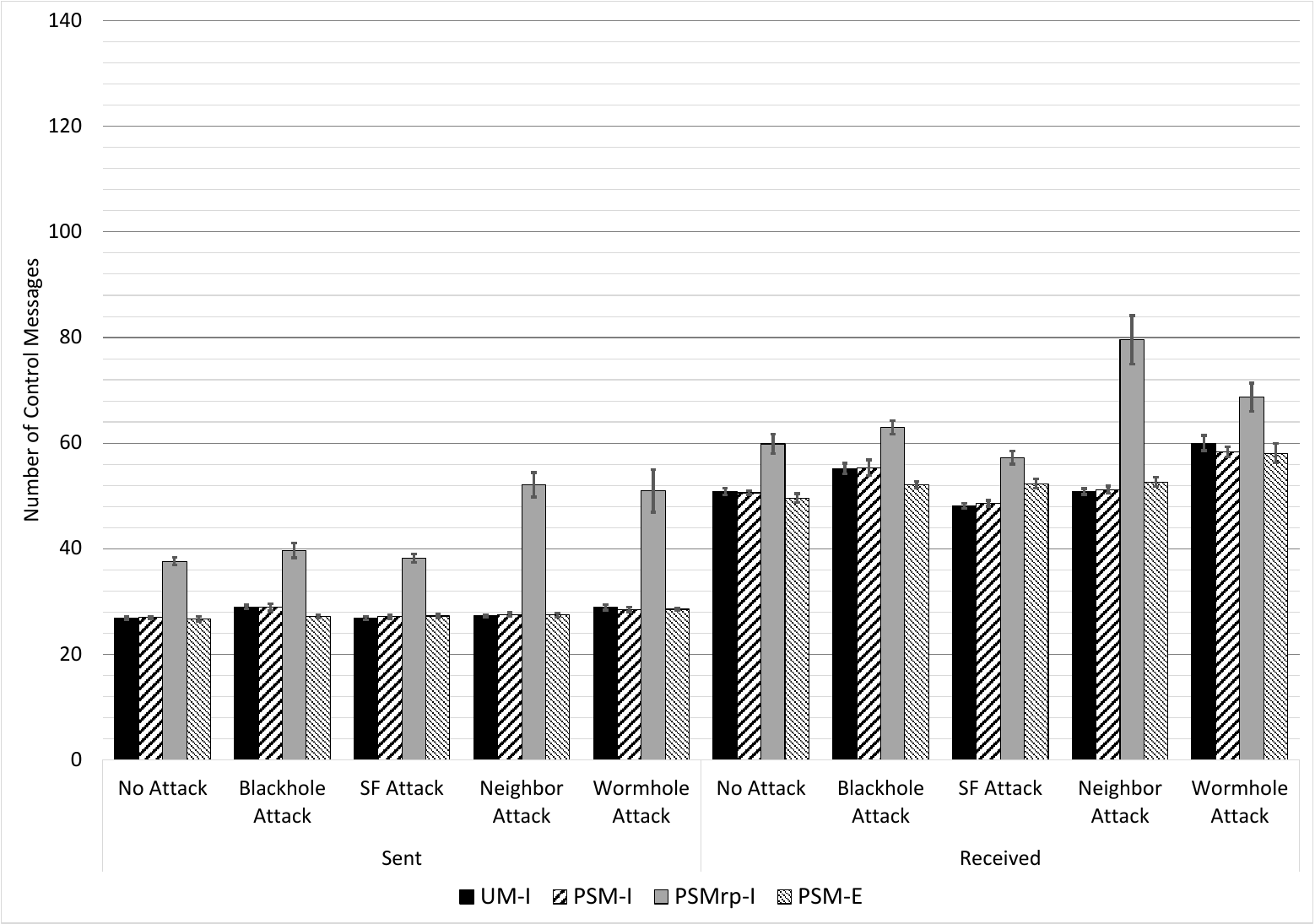}%
		\label{fig_8c}}
	\hfil
	\caption{Simulation results for all four attacks with the first suggestion implemented, using NullRDC RDC protocol.}
	\label{fig_8}
\end{figure*}
\begin{figure*}[!h]
	\centering
	\subfloat[Average packet delivery rate (\gls{pdr}).]{\includegraphics[scale=.29]{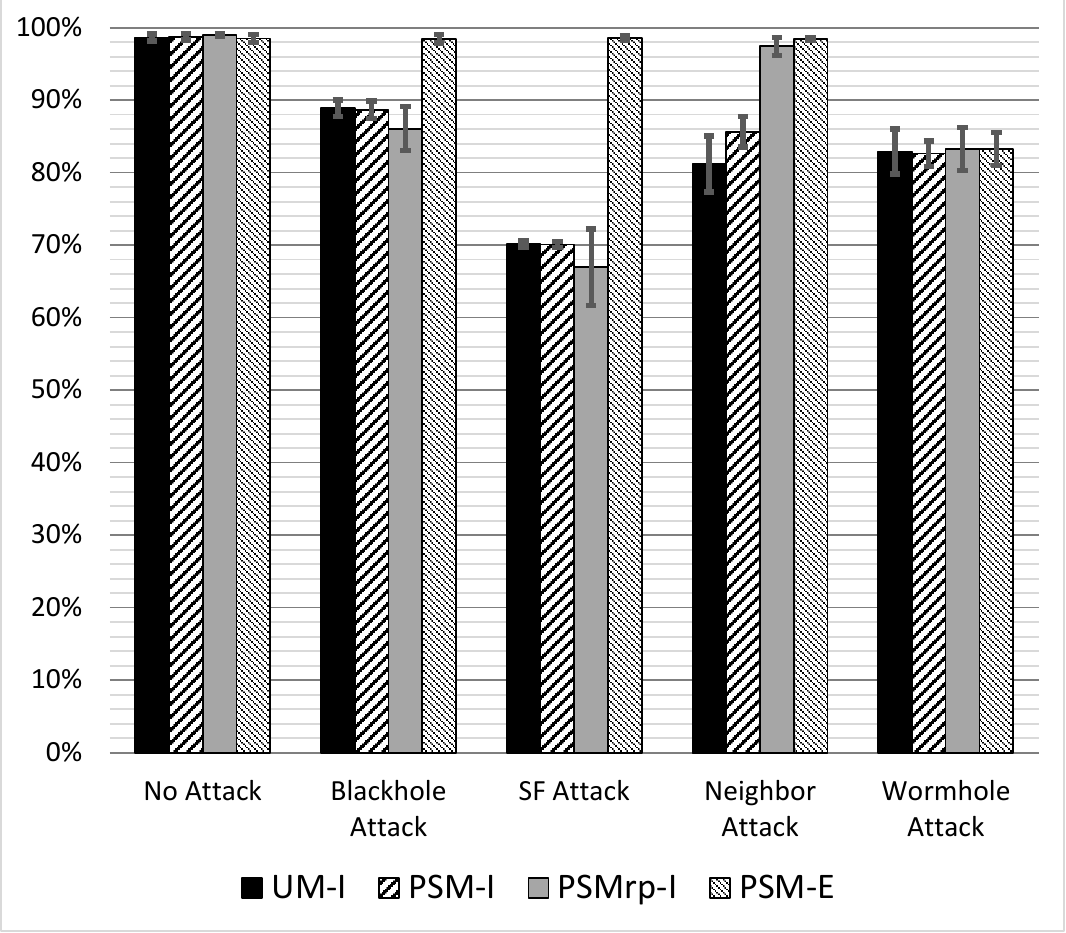}%
		\label{fig_9a}}
	\hfil
	\subfloat[Average network \gls{e2e} latency.]{\includegraphics[scale=.29]{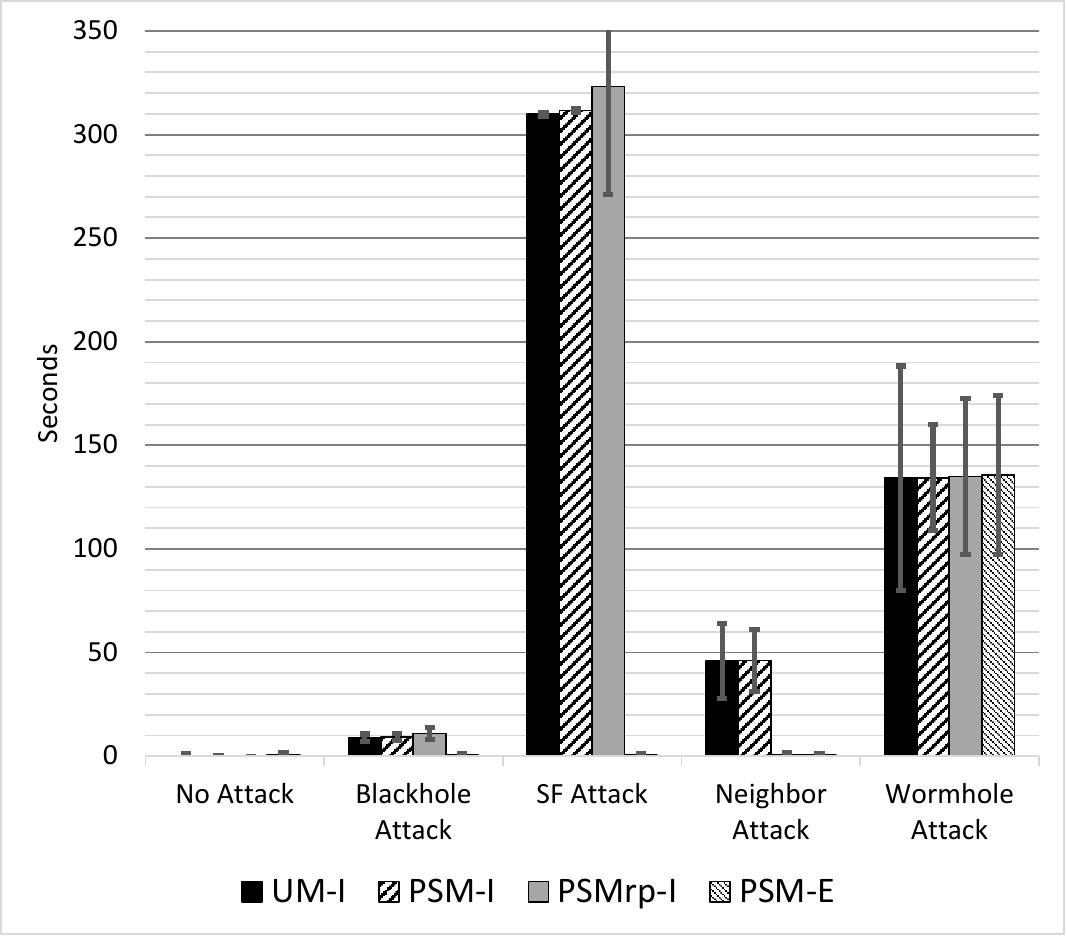}%
		\label{fig_9b}}
	\hfil
	\subfloat[Exchanged \gls{rpl} control messages, per legitimate node.]{\includegraphics[scale=.25]{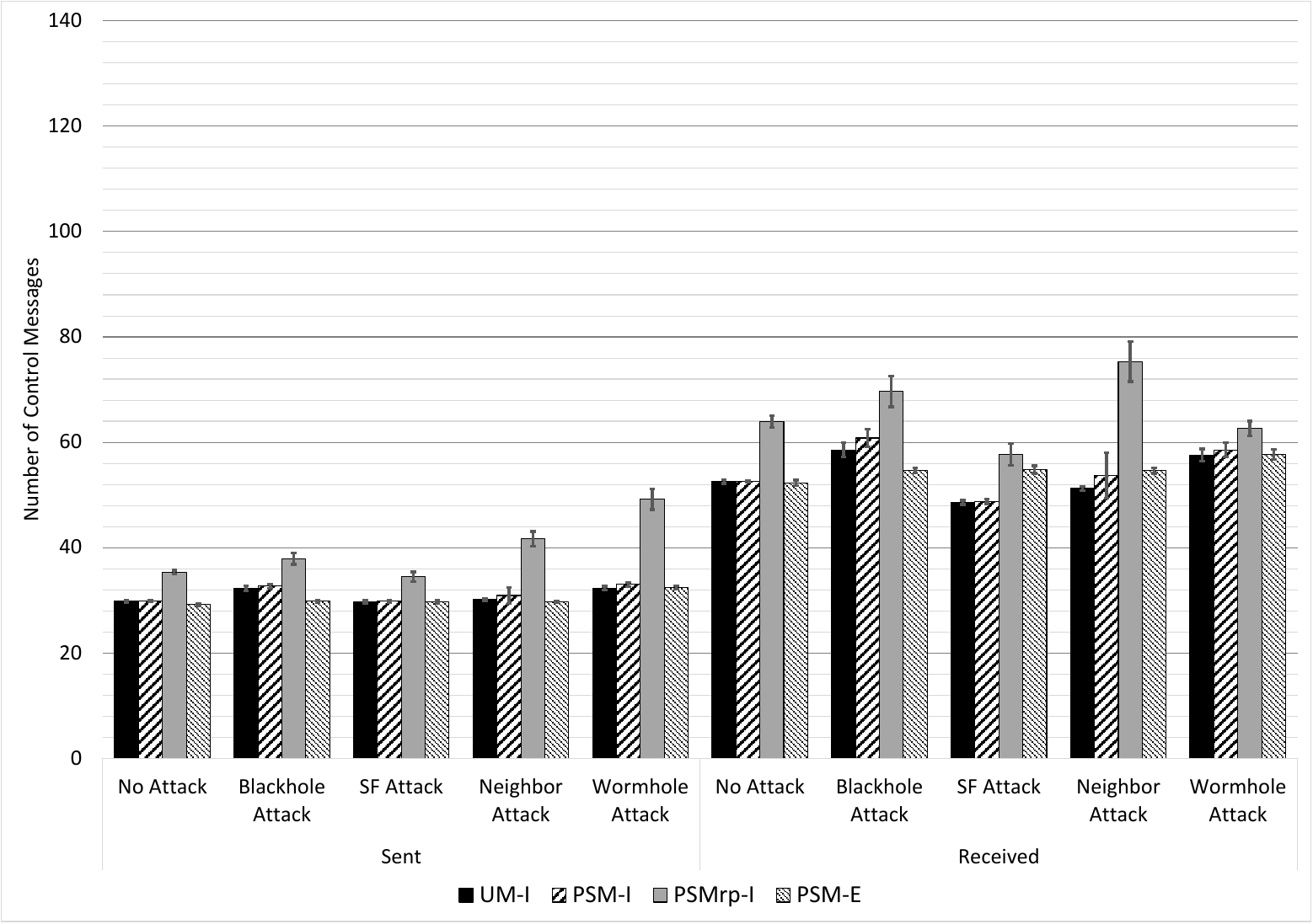}%
		\label{fig_9c}}
	\hfil
	\caption{Simulation results for all four attacks with the second suggestion implemented, using NullRDC RDC protocol.}
	\label{fig_9}
\end{figure*}

\section{Results and Analysis}\label{resultsanalysis}
The results for ContikiMAC and NullRDC sets of experiments are shown in Fig.\ref{fig_2} and Fig.\ref{fig_7}, respectively. Tables \ref{Exp_res_ContikiMAC} and \ref{Exp_res_NullRDC} show the numerical values of all results obtained in this paper. These results are expressed as the average \gls{pdr}, average \gls{e2e} latency, the number of exchanged \gls{rpl} control messages (per legitimate node), and average network power consumption (per received packet). Fig.\ref{fig_3} and Fig. \ref{fig_11} show the routing \gls{dodag} for each scenario that was formed in 90\% of the time in all experiments.

\subsection{ContikiMAC Set Results}
\textbf{Effects on \acrfull{pdr}:} Looking at Fig.\ref{fig_2a}, it is clear that the \gls{rpl} in \gls{psmrpl} successfully mitigated the BH, SF, and NA when the adversary is external with the \gls{pdr} hovering around 98\%.

\begin{figure*}[!ht]
	\centering
	\subfloat[No Attack scenario and SF Attack scenario (all experiments except PSM-E)]{\includegraphics[scale=.256]{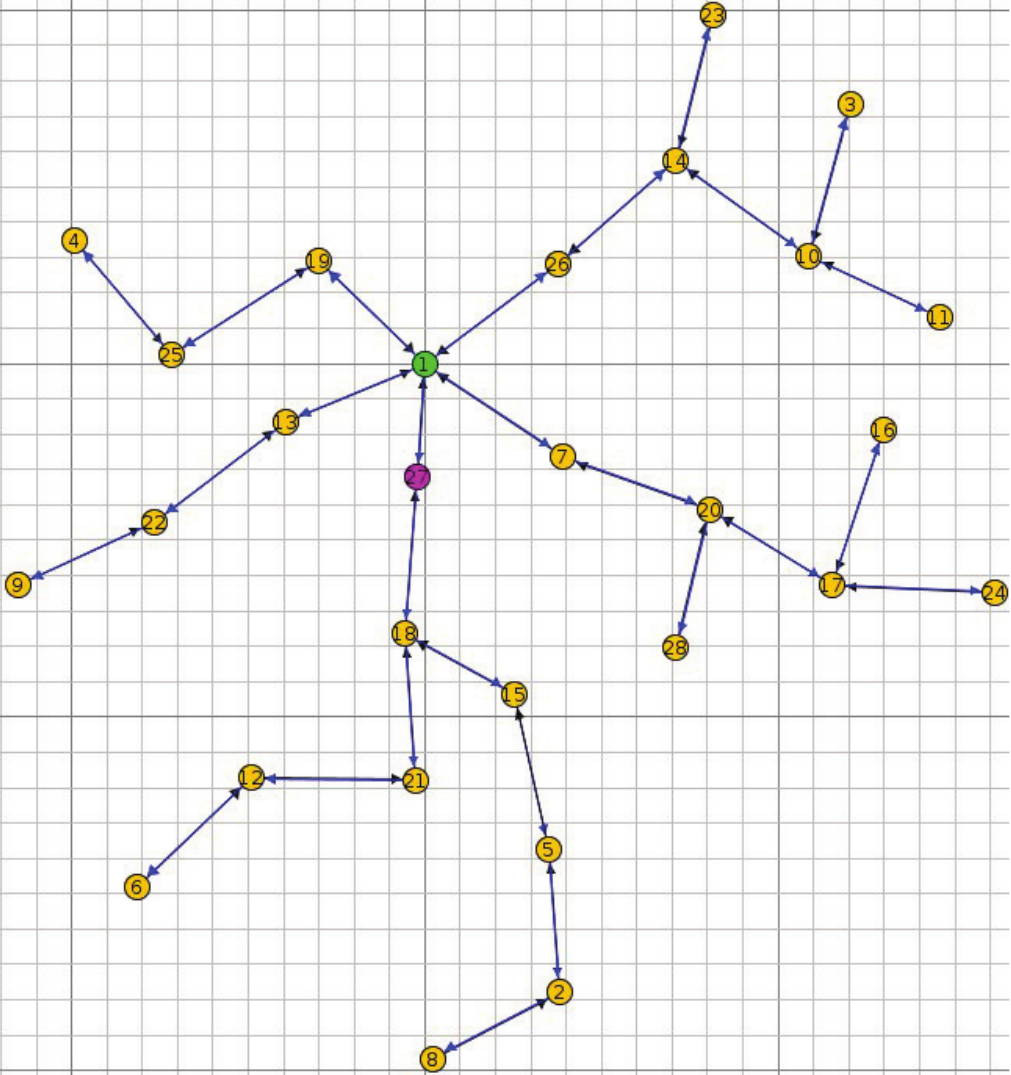}%
		\label{fig_3a}}
	\rulesep
	\subfloat[Blackhole Attack scenario (all experiments), and all scenarios for PSM-E.]{\includegraphics[scale=.25]{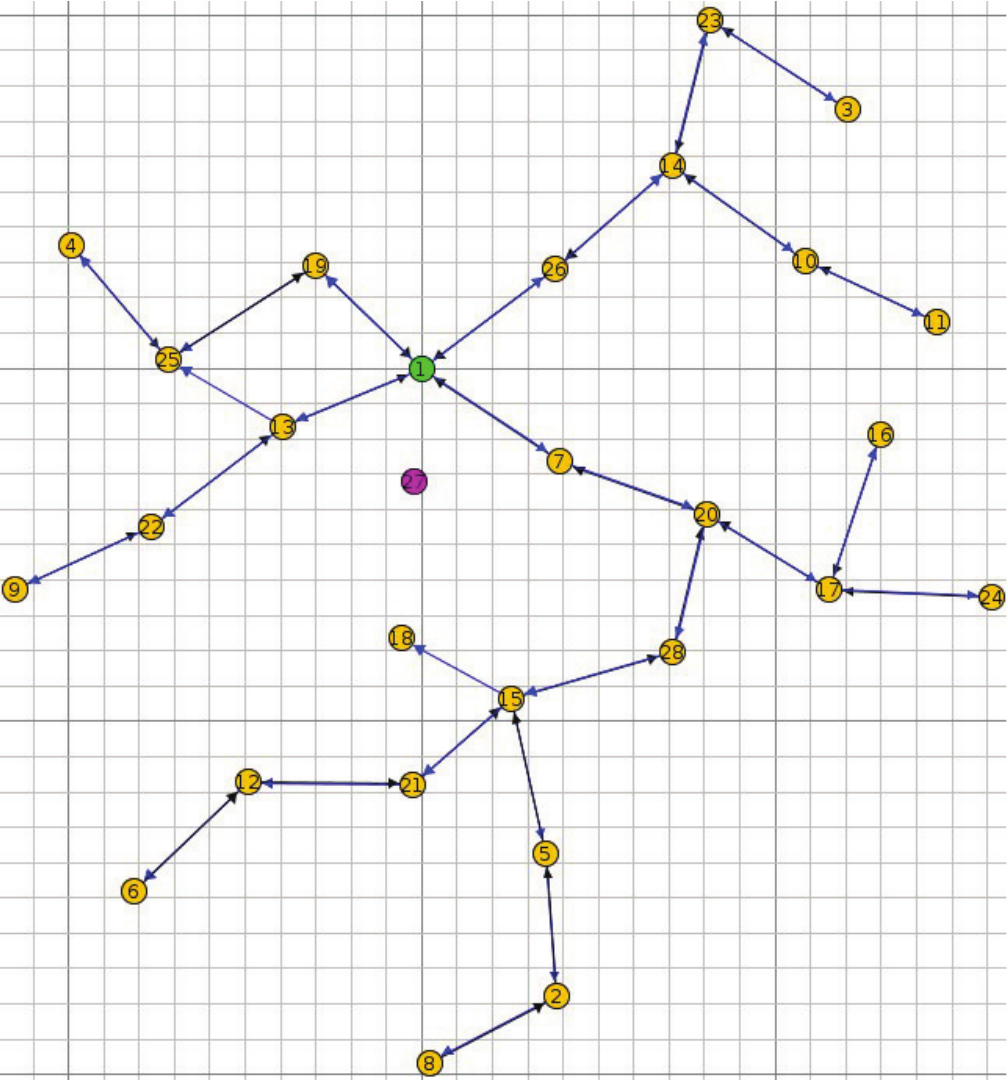}%
		\label{fig_3b}}
	\rulesep
	\subfloat[Neighbor Attack scenario (UM-I and PSM-I).]{\includegraphics[scale=.25]{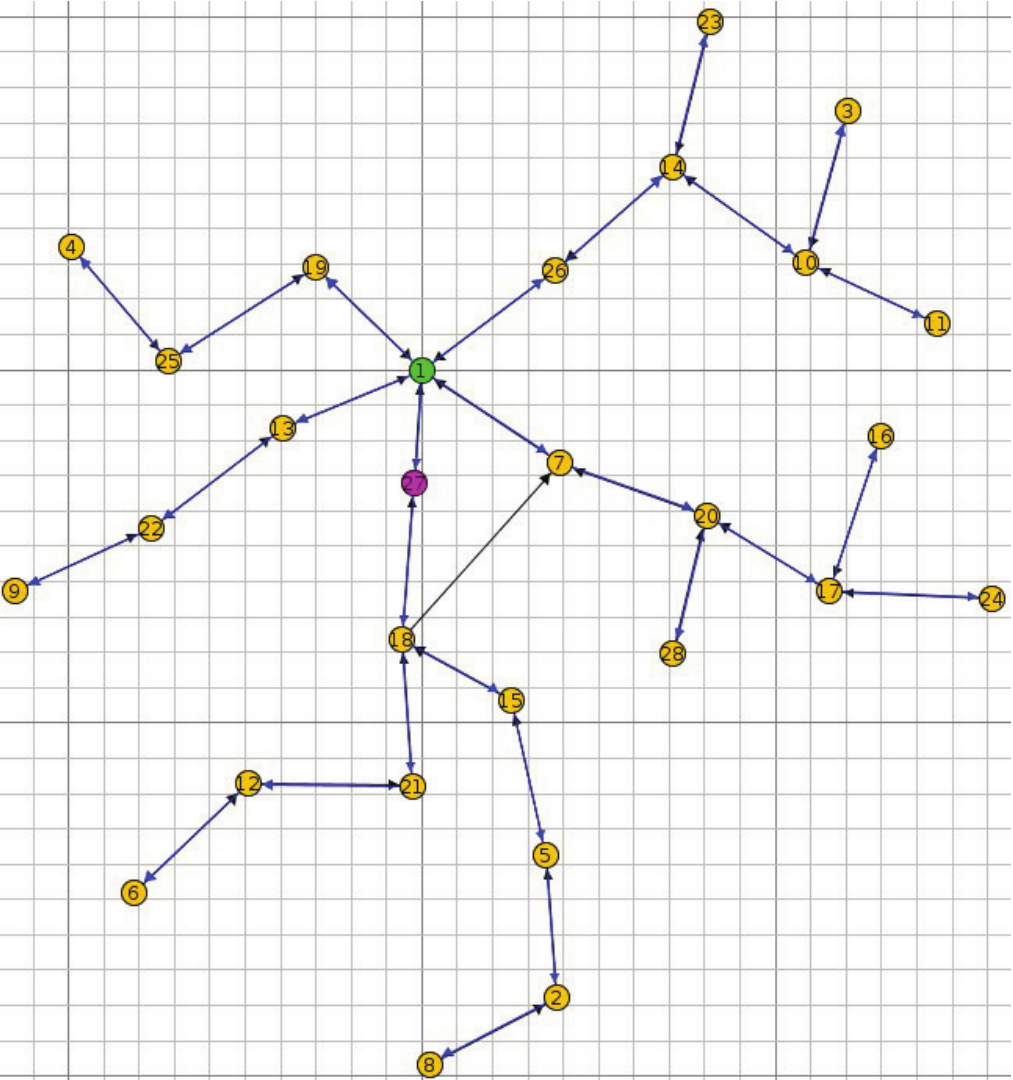}%
		\label{fig_3c}}
	\rulesep
	\subfloat[Neighbor Attack scenario (PSMrp-I).]{\includegraphics[scale=.25]{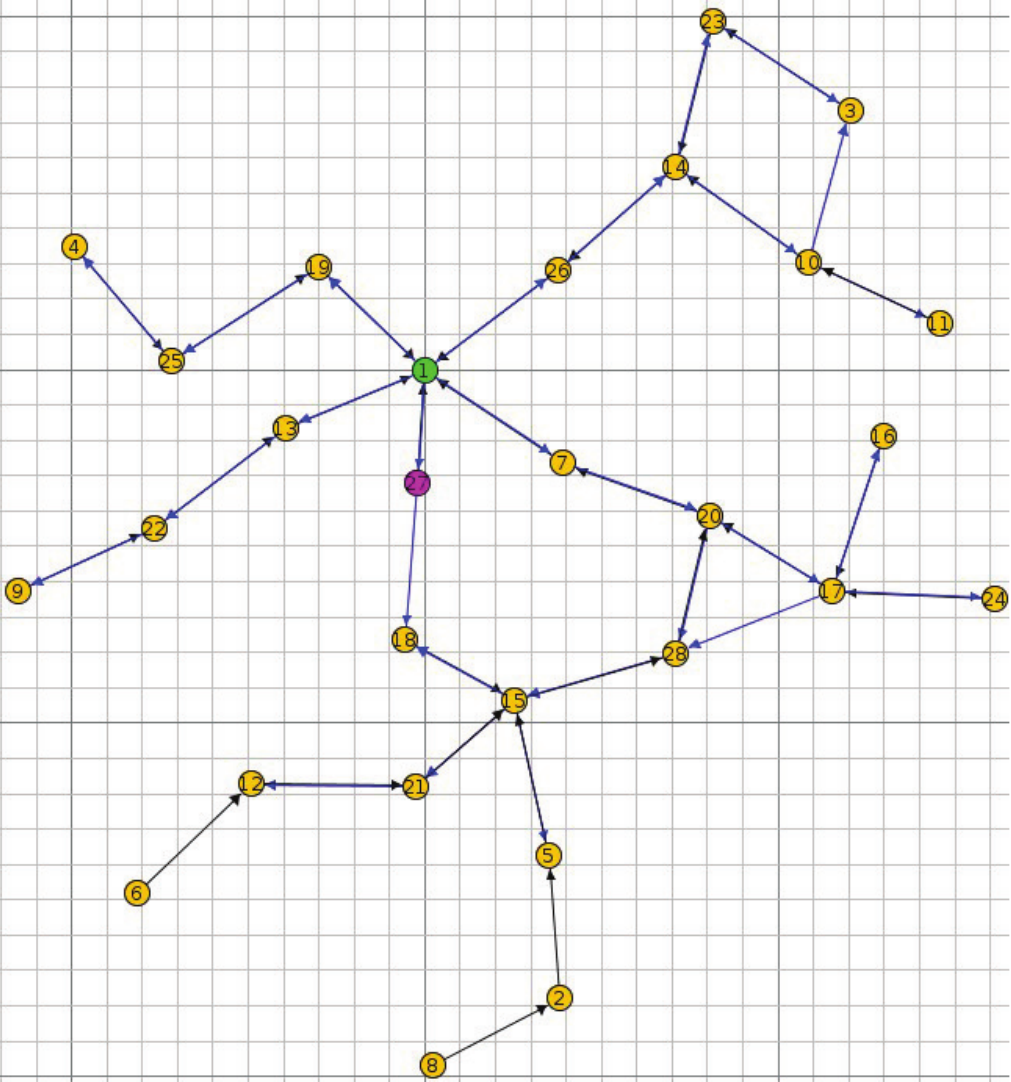}%
		\label{fig_3d}}
	\caption{Routing \glspl{dodag} during the investigated scenarios.}
	\label{fig_3}
\end{figure*}

On the other hand, when the adversary is internal, the SF attack has the most effect (in all experiments) on the \gls{pdr}, decreasing it to a low of 70\%. The main reason behind the success is that the adversary, due to being an active participant in the \gls{dodag} maintenance, is always chosen as the preferred parent for its sub-\gls{dodag}. However, none of their data packets are passed to the sink node. Fig.\ref{fig_3a} shows the routing \gls{dodag} during the SF attack.

For the BH attack, the self-healing mechanisms of \gls{rpl} were always able to detect the unresponsive adversary after approximately ten minutes from the attack launch time (which is the default setting for "dead parent" timeouts in the Contiki \gls{os}) and initiated a local repair for the affected sub-\gls{dodag} to switch to an alternative path. Hence, not all data packets got dropped, which explains why \gls{pdr} is in the range of 80\%. Fig.\ref{fig_3b} shows the routing \gls{dodag} after ten minutes from the BH attack launch time and the isolated adversary.

Finally, for the Neighbor attack, the adversary was able to reduce the \gls{pdr} for the UM-I and PSM-I experiments, as node 18 always chose either node 7 or 13 as its preferred parent (Fig.\ref{fig_3c} shows that node 18 selected node 7 as its preferred parent), due to receiving their \gls{dio} messages through the adversary. Since nodes 7 and 13 are actually out of node 18's range, all packets sent toward them from node 18 and its sub-\gls{dodag} are lost. Hence, the \gls{pdr} is in the same range as in the BH attack scenario. However, activating the replay protection mechanism results in much better \gls{pdr} as the mechanism verifies each \gls{dio} message's original sender before processing its contents. Fig.\ref{fig_3d} demonstrate how the network (in PSMrp-I experiment) opted for the alternative path after a few minutes from launching the NA.

\textbf{Effects on the \gls{e2e} latency:} Confirming our findings mentioned above, Fig.\ref{fig_2b} shows that the \gls{rpl} in \gls{psmrpl} mitigated the BH, SF, and NA when they were launched by an external adversary, keeping the \gls{e2e} latency at a minimum.

Due to the large number of undelivered data packets for the affected nodes, the SF attack had the longest \gls{e2e} latency among all the internal attacks. This effect is, again, due to the adversary's active participation in the \gls{dodag} maintenance.

For the same reason, the BH attack introduced some latency to the network. However, since the affected nodes were able to find an alternative path and were successful in delivering the rest of their data packets, the latency was much lower than in the SF attack scenario.

The situation is more complicated for the NA scenario, as self-healing mechanisms were triggered several times to recover the affected nodes from the attack, which led to even higher \gls{e2e} latency than the BH attack scenario. In general, whenever node 18 switches its preferred parent to node 7 or 13, the sub-\gls{dodag} suffers from Blackhole-like conditions resulting in losing several data packets. In addition, node 18 will either switch its preferred parent back to the adversary when it does not receive \gls{dio} messages from the "ghost parent" (node 7 or node 13), or initiate a local repair procedure (if \gls{dodag} inconsistencies were detected) that results in the whole sub-\gls{dodag} choosing the alternative path to deliver their packets. Either way, it will add more latency to the network.
Using the replay protection will significantly reduce the latency from the NA, as node 18 will not switch its preferred parent as long as it does not receive the correct \gls{cc} response from nodes 7 and 13.

\textbf{Effects on the exchanged number of \gls{rpl}'s control messages:} As seen in Fig.\ref{fig_2c}, the number of control messages exchanged in the network is almost the same for all experiments and all the scenarios, with the replay protection mechanisms adding a bit more control messages. The exception of this conclusion is the NA scenario with \gls{rpl} in \gls{psmrpl}rp. In this particular case, the replay protection mechanism introduced a much higher number of control messages, due to the exchange of the \gls{cc} messages whenever a "ghost" \gls{dio} message is received by nodes 7, 13, or 18.

It is worth noting that the number of received control messages is always higher than the sent one because many of the sent control messages are multicast messages which will be received by all neighboring nodes of the sender.

\textbf{Effects on power consumption:} Fig.\ref{fig_2d} shows the average network power consumption per received packet, as it gives a more accurate look into the effect of the attacks on the power consumption than just using the regular average power consumption readings.

Looking at the results of the external adversary experiment in the No Attack scenario, we can see that the power consumption is a bit higher than the same scenario in the other experiments. The reason is that the data packets from the affected nodes are taking the alternative and longer path, i.e., more power is used by the nodes on that path. However, the power consumption pattern is identical in all the scenarios of the external adversary experiment, which indicates no effect from the attacks; hence, successful mitigation of the attacks.
\begin{table*}[!ht]
	\renewcommand{\arraystretch}{1.3}
	\centering
	\caption{Simulation results for the four experiments (three attacks scenarios), using ContikiMAC RDC protocol.}
	\resizebox{\textwidth}{!}{
		\begin{tabular}{l*{20}{>{\centering\arraybackslash}m{0.06\textwidth}}}
			\toprule
			\multicolumn{1}{c}{\multirow{3}{*}{\textbf{Scenario}}} & \multicolumn{4}{c}{\multirow{2}{*}{\textbf{Average PDR}}} & \multicolumn{4}{c}{\multirow{2}{*}{\textbf{Average E2E Latency (in seconds)}}} & \multicolumn{4}{c}{\multirow{2}{4cm}{\centering\textbf{Average Power Consumption \newline (per received packet - in mWatt)}}} & \multicolumn{8}{c}{\textbf{Exchanged RPL Control Messages (per legitimate node)}} \\
			\cmidrule{14-21}          & \multicolumn{4}{c}{}         & \multicolumn{4}{c}{}         & \multicolumn{4}{c}{}         & \multicolumn{4}{c}{\textbf{Sent}}     & \multicolumn{4}{c}{\textbf{Received}} \\
			\cmidrule{2-21}          & \textbf{UM-I}  & \textbf{PSM-I} & \textbf{PSMrp-I} & \textbf{PSM-E} & \textbf{UM-I}  & \textbf{PSM-I} & \textbf{PSMrp-I} & \textbf{PSM-E} & \textbf{UM-I}  & \textbf{PSM-I} & \textbf{PSMrp-I} & \textbf{PSM-E} & \textbf{UM-I}  & \textbf{PSM-I} & \textbf{PSMrp-I} & \textbf{PSM-E} & \textbf{UM-I}  & \textbf{PSM-I} & \textbf{PSMrp-I} & \textbf{PSM-E} \\
			\midrule
			\multicolumn{21}{c}{Default Setting - (Fig.\ref{fig_2})}\\
			\midrule
			\textbf{No Attack} & 98.40\% & 98.36\% & 98.10\% & 97.45\% & 0.480 & 0.374 & 0.575 & 0.656 & 0.00664 & 0.00658 & 0.00715 & 0.00809 & 29    & 29    & 32    & 30    & 63    & 60    & 55    & 68 \\
			\midrule
			\textbf{BH Attack} & 80.55\% & 80.21\% & 80.31\% & 97.82\% & 23.792 & 27.307 & 28.277 & 0.672 & 0.00931 & 0.00900 & 0.00951 & 0.00813 & 34    & 33    & 36    & 30    & 78    & 73    & 78    & 71 \\
			\midrule
			\textbf{SF Attack} & 69.56\% & 69.85\% & 69.63\% & 97.11\% & 324.754 & 322.885 & 320.194 & 0.725 & 0.00920 & 0.00927 & 0.00975 & 0.00814 & 30    & 30    & 37    & 31    & 66    & 65    & 67    & 72 \\
			\midrule
			\textbf{NA Attack} & 79.49\% & 83.43\% & 92.72\% & 96.97\% & 45.506 & 49.610 & 8.885 & 0.680 & 0.00809 & 0.00793 & 0.01035 & 0.00804 & 32    & 32    & 52    & 32    & 70    & 70    & 127   & 73 \\
			\midrule
			\multicolumn{21}{c}{First Suggestion - (Fig.\ref{fig_5})}\\
			\midrule
			\textbf{No Attack} & 98.04\% & 98.08\% & 98.19\% & 97.55\% & 0.806 & 0.529 & 0.533 & 1.038 & 0.00650 & 0.00670 & 0.00696 & 0.00732 & 31    & 31    & 34    & 32    & 74    & 76    & 81    & 86 \\
			\midrule
			\textbf{BH Attack} & 85.32\% & 84.94\% & 85.50\% & 97.93\% & 17.783 & 17.998 & 17.695 & 0.562 & 0.00781 & 0.00804 & 0.00824 & 0.00733 & 33    & 34    & 37    & 32    & 83    & 85    & 90    & 85 \\
			\midrule
			\textbf{SF Attack} & 77.82\% & 75.79\% & 76.92\% & 98.04\% & 213.223 & 231.112 & 214.871 & 0.485 & 0.00779 & 0.00825 & 0.00885 & 0.00724 & 31    & 31    & 36    & 32    & 75    & 76    & 85    & 85 \\
			\midrule
			\textbf{NA Attack} & 86.00\% & 86.50\% & 94.74\% & 97.70\% & 44.424 & 33.717 & 4.486 & 0.631 & 0.00716 & 0.00708 & 0.00930 & 0.00714 & 32    & 32    & 51    & 32    & 77    & 76    & 132   & 81 \\
			\midrule
			\multicolumn{21}{c}{Second Suggestion - (Fig.\ref{fig_6})}\\
			\midrule
			\textbf{No Attack} & 98.32\% & 97.71\% & 98.02\% & 97.02\% & 0.425 & 1.076 & 0.889 & 1.990 & 0.00696 & 0.00750 & 0.00741 & 0.00895 & 29    & 33    & 36    & 34    & 53    & 59    & 60    & 61 \\
			\midrule
			\textbf{BH Attack} & 88.78\% & 87.81\% & 87.74\% & 97.08\% & 8.213 & 10.367 & 10.591 & 1.896 & 0.00892 & 0.00922 & 0.00968 & 0.00878 & 35    & 36    & 41    & 34    & 66    & 67    & 71    & 61 \\
			\midrule
			\textbf{SF Attack} & 69.41\% & 69.33\% & 69.59\% & 97.02\% & 312.429 & 312.945 & 307.448 & 1.843 & 0.00993 & 0.00974 & 0.01037 & 0.00864 & 33    & 32    & 37    & 33    & 55    & 53    & 57    & 61 \\
			\midrule
			\textbf{NA Attack} & 79.93\% & 83.05\% & 93.84\% & 96.74\% & 52.318 & 48.455 & 4.596 & 1.960 & 0.00888 & 0.00847 & 0.01124 & 0.00895 & 35    & 34    & 58    & 34    & 61    & 59    & 104   & 62 \\
			\bottomrule
		\end{tabular}%
	}
	\label{Exp_res_ContikiMAC}%
\end{table*}%
\begin{table*}[!ht]
	\renewcommand{\arraystretch}{1.3}
	\centering
	\caption{Simulation results for the four experiments (four attacks scenarios), using NullRDC RDC protocol.}
	\resizebox{\textwidth}{!}{
		\begin{tabular}{l*{16}{>{\centering\arraybackslash}m{0.06\textwidth}}}
			\toprule
			\multicolumn{1}{c}{\multirow{3}{*}{\textbf{Scenario}}} & \multicolumn{4}{c}{\multirow{2}{*}{\textbf{Average PDR}}} & \multicolumn{4}{c}{\multirow{2}{*}{\textbf{Average E2E Latency (in seconds)}}} & \multicolumn{8}{c}{\textbf{Exchanged RPL Control Messages (per legitimate node)}} \\
			\cmidrule{10-17}          & \multicolumn{4}{c}{}         & \multicolumn{4}{c}{}         & \multicolumn{4}{c}{\textbf{Sent}}     & \multicolumn{4}{c}{\textbf{Received}} \\
			\cmidrule{2-17}          & \textbf{UM-I}  & \textbf{PSM-I} & \textbf{PSMrp-I} & \textbf{PSM-E} & \textbf{UM-I}  & \textbf{PSM-I} & \textbf{PSMrp-I} & \textbf{PSM-E} & \textbf{UM-I}  & \textbf{PSM-I} & \textbf{PSMrp-I} & \textbf{PSM-E} & \textbf{UM-I}  & \textbf{PSM-I} & \textbf{PSMrp-I} & \textbf{PSM-E} \\
			\midrule
			\multicolumn{17}{c}{Default Setting - (Fig.\ref{fig_7})}\\
			\midrule
			\textbf{No Attack} & 99.15\% & 98.74\% & 98.22\% & 98.80\% & 0.033 & 0.034 & 0.047 & 0.036 & 27    & 27    & 36    & 26    & 47    & 47    & 53    & 45 \\
			\midrule
			\textbf{BH Attack} & 80.28\% & 80.44\% & 80.72\% & 98.93\% & 31.302 & 28.266 & 26.500 & 0.036 & 30    & 30    & 39    & 27    & 54    & 54    & 58    & 47 \\
			\midrule
			\textbf{SF Attack} & 70.07\% & 70.16\% & 69.83\% & 98.52\% & 323.857 & 323.858 & 329.444 & 0.036 & 28    & 28    & 37    & 27    & 49    & 50    & 54    & 47 \\
			\midrule
			\textbf{NA Attack} & 79.57\% & 82.77\% & 94.98\% & 98.60\% & 56.762 & 53.242 & 2.923 & 0.036 & 28    & 28    & 57    & 27    & 49    & 49    & 86    & 47 \\
			\midrule
			\textbf{WH Attack} & 78.32\% & 78.46\% & 82.59\% & 79.19\% & 218.413 & 223.518 & 159.881 & 207.210 & 29    & 28    & 47    & 29    & 52    & 51    & 58    & 51 \\
			\midrule
			\multicolumn{17}{c}{First Suggestion - (Fig.\ref{fig_8})}\\
			\midrule
			\textbf{No Attack} & 98.84\% & 98.69\% & 98.02\% & 98.89\% & 0.052 & 0.038 & 0.052 & 0.040 & 27    & 27    & 38    & 27    & 51    & 51    & 60    & 50 \\
			\midrule
			\textbf{BH Attack} & 82.61\% & 84.64\% & 83.86\% & 98.78\% & 23.728 & 19.043 & 16.321 & 0.036 & 29    & 29    & 40    & 27    & 55    & 55    & 63    & 52 \\
			\midrule
			\textbf{SF Attack} & 73.73\% & 72.93\% & 73.44\% & 98.48\% & 284.978 & 291.616 & 277.439 & 0.036 & 27    & 27    & 38    & 27    & 48    & 49    & 57    & 52 \\
			\midrule
			\textbf{NA Attack} & 85.24\% & 85.49\% & 97.24\% & 98.86\% & 41.850 & 47.155 & 0.566 & 0.036 & 27    & 28    & 52    & 27    & 51    & 51    & 80    & 53 \\
			\midrule
			\textbf{WH Attack} & 80.08\% & 80.74\% & 85.07\% & 80.99\% & 199.385 & 200.145 & 135.046 & 190.021 & 29    & 28    & 51    & 29    & 60    & 58    & 69    & 58 \\
			\midrule
			\multicolumn{17}{c}{Second Suggestion - (Fig.\ref{fig_9})}\\
			\midrule
			\textbf{No Attack} & 98.66\% & 98.72\% & 98.97\% & 98.50\% & 0.362 & 0.164 & 0.036 & 0.691 & 30    & 30    & 35    & 29    & 53    & 53    & 64    & 52 \\
			\midrule
			\textbf{BH Attack} & 88.90\% & 88.68\% & 86.09\% & 98.46\% & 8.976 & 9.161 & 10.962 & 0.479 & 32    & 33    & 38    & 30    & 59    & 61    & 70    & 55 \\
			\midrule
			\textbf{SF Attack} & 70.18\% & 70.11\% & 66.94\% & 98.58\% & 309.822 & 311.586 & 323.208 & 0.426 & 30    & 30    & 34    & 30    & 49    & 49    & 58    & 55 \\
			\midrule
			\textbf{NA Attack} & 81.19\% & 85.66\% & 97.45\% & 98.42\% & 45.938 & 46.157 & 0.823 & 0.503 & 30    & 31    & 42    & 30    & 51    & 54    & 75    & 55 \\
			\midrule
			\textbf{WH Attack} & 82.91\% & 82.62\% & 83.27\% & 83.25\% & 134.065 & 134.374 & 135.079 & 135.804 & 32    & 33    & 49    & 32    & 58    & 59    & 63    & 58 \\
			\bottomrule
		\end{tabular}%
	}
	\label{Exp_res_NullRDC}%
\end{table*}%

For all internal-adversary experiments, the power consumption patterns (per scenario) are very similar between \gls{rpl} in \acrshort{umrpl} and \gls{psmrpl} for the No Attack, BH, and SF attacks scenarios, with the replay protection mechanism having a bit more power consumption than the rest. This is because many data packets were not delivered, and the power consumed for their unsuccessful deliveries is entirely wasted.

Now, it is clear from Fig.\ref{fig_2d} that using the replay protection significantly increases the average power consumption when the NA is launched, even if almost all of the sent data packets were delivered successfully. This time, the reason behind this behavior is the increased number of control messages exchanged to mitigate the attack, as seen in Fig.\ref{fig_2c}.

\subsection{NullRDC Set Results}
\begin{figure}[!t]
	\centering
	\includegraphics[height=45mm, width=50mm]{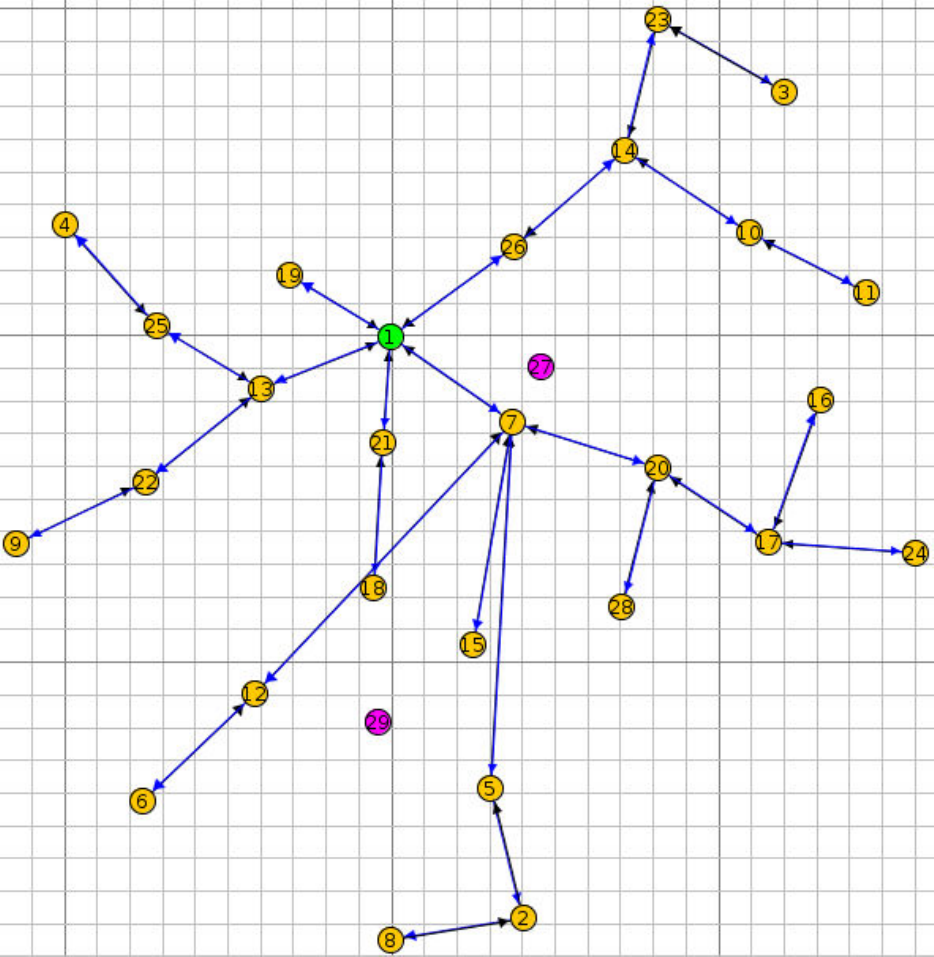}
	\caption{Routing \gls{dodag} during Wormhole Attack scenario (all experiments)}
	\label{fig_11}
\end{figure}

Comparing Fig.\ref{fig_7} to Fig.\ref{fig_2}, it is clear that we have similar results for the first four scenarios (No Attack, BH, SF, and NA) in both \gls{rdc} protocols, i.e., NullRDC and ContikiMAC. Hence, the focus of this analysis will be on the Wormhole (WH) attack scenario. The effects of the WH attack on the routing \gls{dodag} can be seen in Fig.\ref{fig_11}.

\textbf{Effects on \acrfull{pdr}:} The WH attack successfully lowered the \gls{pdr} to the low 80th percentile in all scenarios, regardless of the used \gls{rpl}'s secure mode or the adversary type. Our observation shows that the reason behind such behavior is because the adversaries are transparent to the network and that all control messages (from both sides of the wormhole) are forwarded and received within their time-windows, deceiving the legitimate nodes to think they are in close proximity.

\textbf{Effects on the \gls{e2e} latency:} Since most of the affected nodes were unable to deliver their data packets successfully, the average \gls{e2e} latency of the network rose to 200 seconds - see Fig.\ref{fig_7b}. \gls{rpl}'s replay protection mechanism slightly reduced the effect of the WH attack. However, this is due to having slight delays with the \gls{cc} message exchanges.

\textbf{Effects on the exchanged number of \gls{rpl}'s control messages:} At a first look, it is evident that using \gls{rpl} over NullRDC protocol reduces the number of exchanged control packets compared to using the ContikiMAC protocol, which has been documented in \cite{Barnawi2019}. Besides that, the WH attack exchanged a similar number of control messages as in the other attacks, with the replay protection mechanism in \gls{psmrpl}rp slightly increasing that number over the other experiments.

\section{Discussions}\label{discussionSec}
Based on the analysis of the obtained results, we can put the following observations and, as a result, some suggestions to improve \gls{rpl}'s response to the investigated attacks.

\subsection{Observations}\label{observ}
\begin{itemize}
	\item Using \gls{rpl} in \gls{psmrpl} (and by extension, the \acrshort{asmrpl}) can mitigate the external adversaries of the Blackhole, Selective-Forward, and Neighbor attacks, as long as the adversary does not run \gls{rpl} in any secure mode.
	\item \gls{rpl}'s performance using \gls{psmrpl} (without the replay protection mechanism) is similar to that when using \acrshort{umrpl}, but with the added benefit of mitigating the external adversaries of the BH, SF, and Neighbor attacks as investigated in this paper.
	\item \gls{rpl}'s secure modes cannot mitigate out-of-band Wormhole attacks (with the NullRDC protocol at the Link layer) as their adversaries can operate external to the network.
	\item It is worth mentioning that we ran another experiment (using ContikiMAC) that had the external adversary running \gls{rpl} in \gls{psmrpl} while not knowing the encryption key used by the legitimate nodes. The results from that experiment were identical to the PSM-E experiment except for the Neighbor attack scenario, which was successfully launched. Since each type of \gls{rpl} control messages has its unique \gls{icmp} "Code" value, with the secure versions having different values than the unsecure ones, only a node that runs \gls{rpl} in \gls{psmrpl}/\acrshort{asmrpl} could identify the secure versions of \gls{rpl} control messages. Hence, the adversary was able to identify \gls{rpl}'s secure \gls{dio} messages and replay them.
	\item Enabling \gls{rpl}'s replay protection mechanism will significantly reduce the effect of Neighbor attacks on \gls{pdr} and \gls{e2e} latency. However, in its current implementation, it will increase the power consumption as well, which can lead to energy depletion of the devices. In theory, an adversary can replay \gls{dio} messages regularly to keep the affected nodes always busy with the consistency checks, leading to depletion of their energy and shutdown.
	\item \gls{rpl}'s secure modes require more memory and storage spaces than the unsecured mode, which means not all \gls{iot} devices can use them -- see \S\ref{EvalSetup}.
\end{itemize}

\subsection{Suggestions to Reduce the Effects of Routing Attacks on RPL's Performance}\label{Recomm}
Based on the observations mentioned above, we propose the following suggestions to help reduce the effects of routing attacks on \gls{rpl}'s performance, without introducing any additional security mechanisms or systems.
\begin{enumerate}
	\item Designing the network topology in a way where there are more alternative paths toward the root node and more neighbors per node. This would decrease the recovery time required for nodes to overcome a Blackhole attack and reduce the effects from the other investigated attacks on \gls{pdr} and \gls{e2e} latency.
	\item Reducing the timeout duration after which an \gls{rpl} router should declare a preferred parent as "dead". Currently, ContikiRPL uses fixed timeout values for the upward (UIP\_CONF\_ND6\_REACHABLE\_TIME) and downward routes (RPL\_CONF\_DEFAULT\_LIFETIME), both set to 10 minutes. Reducing these values could decrease the \gls{e2e} latency and increase the \gls{pdr} of the network under some attack situations. However, static decrements may also increase power consumption when there are no attacks. Our recommendation is to use a dynamic approach for adapting these timeout values to the network's changing conditions. For example, randomizing the timeout values after each expiration, or using the \acrlong{6lowpan}-Neighbor Discovery (\acrshort{6lowpan}-ND) protocol\cite{RFC6775,RFC8505,ietf-6lo-ap-nd-12}, which aids \gls{rpl} to detect node's neighbors and checks their status in a resource-friendly way.
\end{enumerate}

\subsection{Evaluation of the Proposed Suggestions}
\begin{figure}[!t]
	\centering
	\includegraphics[scale=0.45]{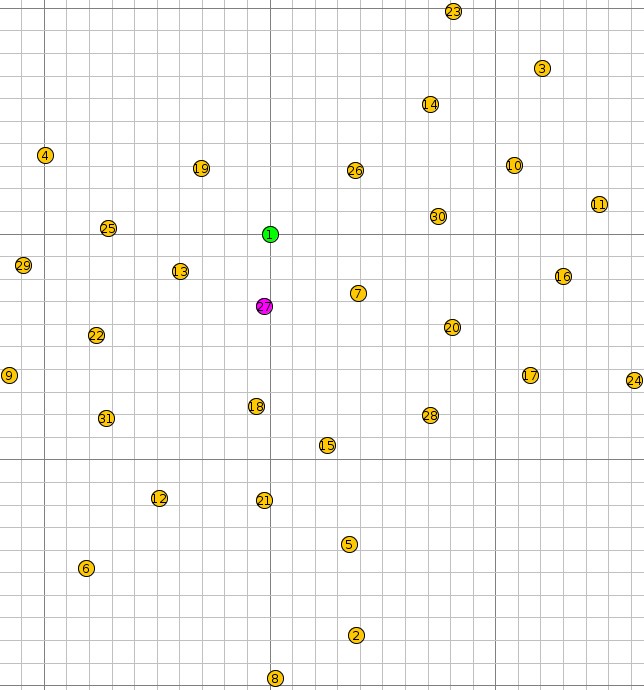}
	\caption{Network topology for the first suggestion.}
	\label{fig_4}
\end{figure}

For the first suggestion, having more routes toward the root node means adding more routing nodes. Hence, we added three routing nodes (29, 30, and 31) to the topology - see Fig.\ref{fig_4}. In the Wormhole attack scenario, we also added three routing nodes (30, 31, and 32), which are located at the same positions as in Fig.\ref{fig_4} but with the topology in Fig.\ref{fig_10}.

To evaluate the effect of the second suggestion on \gls{rpl}'s performance under the investigated attacks,  both "dead parent" timeouts (see \S\ref{Recomm}) were set to five minutes. The use of a fixed value instead of a dynamic approach was used to examine the effect of the reduced "dead parent" timeouts only. The topology for the evaluation is the same topology used in Fig.\ref{fig_1}.

The whole evaluation was conducted using the same metrics and methodology as in \S\ref{RPLEval}.

\textbf{Effects on \acrfull{pdr}:} Comparing Fig.\ref{fig_5a} to Fig.\ref{fig_2a} (ContikiMAC) and Fig.\ref{fig_8a} to Fig.\ref{fig_7a} (NullRDC), we can see that the first suggestion slightly enhanced the network's \acrshort{pdr} for the BH, SF, and NA scenarios, adding about 6\% more delivered packets. From our observation, the reason behind this improvement is that some of the affected nodes chose the new alternative routes, minimizing the effect of the investigated attacks.

On the other hand, the second suggestion affected only the BH scenario, increasing the \gls{pdr} to a respected 88\% - this is clear from comparing Fig.\ref{fig_6a} to Fig.\ref{fig_2a} (ContikiMAC) and Fig.\ref{fig_9a} to Fig.\ref{fig_7a} (NullRDC). The reduced timeouts caused the effected nodes to detect the adversary parent faster and switch to a different parent. However, this suggestion does not have any effect in the case of the other attacks, since their adversary reacts to received messages, unlike the Blackhole's adversary.

However, neither suggestion had any effect on \gls{rpl}'s performance in the Wormhole attack scenario.

\textbf{Effects on the \gls{e2e} latency:} Fig.\ref{fig_5b} and \ref{fig_8b} show that the first suggestion decreased the \gls{e2e} in the case of the SF scenario, especially for the ContikiMAC protocol ($\sim$220 seconds, down from $\sim$320 seconds) compared to NullRDC ($\sim$270 seconds, down from $\sim330$ seconds). Again, this is because some of the affected nodes chose the alternative routes away from the adversary and more data packets are delivered.

As for the second suggestion (Fig.\ref{fig_6b} and \ref{fig_9b}), the main enhancement occurred is in the case of the BH scenario (10 seconds down from 30 seconds). As the affected nodes were able to detect the dead adversary parent much faster, the total \gls{e2e} was reduced by more than 50\%.

\textbf{Effects on the exchanged number of \gls{rpl}'s control messages:} As seen in Fig.\ref{fig_5d}, The first suggestion increased the number of received control messages for the ContikiMAC set. However, the reason this time is the added routing nodes themselves but not due to the attacks. On the other hand, the second suggestion (see Fig.\ref{fig_6d}) slightly reduced the number of received \gls{rpl} control messages, especially in the NA scenario.

From Fig.\ref{fig_8c} and \ref{fig_9c}, we can see that both suggestions do not have any effect on the exchanged control messages when NullRDC is used. This is due to the always-on radio and the simpler sending mechanism. 

\textbf{Effects on power consumption:} Figures \ref{fig_5c} and \ref{fig_6c} show that the average network power consumption (per received packet) has been reduced for the first suggestion while increased for the second one. The reason behind the reduction for the first suggestion is that more data packets are delivered successfully. However, the power consumption increase in the second suggestion experiments is because more probing is performed for parent's freshness (due to the shorter timeouts). It is worth mentioning that this analysis is only valid for ContikiMAC set and not NullRDC, as we were not able to collect usable power readings for the latter - see \S\ref{ImptCh}.

From the discussion above, we can conclude that, individually, our two suggestions have mostly a positive effect on the network when under an attack. Hence, combining both of the suggestions with a dynamic timeout setup would further enhance \gls{rpl}'s performance without taxing the scarce resources of the nodes. This, however, is still being investigated.

\section{Conclusion}\label{Conc}
In this paper, we evaluated the performance of \gls{rpl} and its security mechanisms under the presence of four common routing attacks (the Blackhole, Selective-Forward, Neighbor, and Wormhole attacks). The evaluation was carried using two widely used \gls{rdc} protocols, ContikiMAC and NullRDC. Our analysis showed that using \gls{rpl} in \gls{psmrpl} can mitigate external adversaries of the investigated attacks (except for the Wormhole attack) as long as the adversaries do not run \gls{rpl} in \gls{psmrpl}/\acrshort{asmrpl}. It also showed that using \gls{rpl} in \gls{psmrpl}/\acrshort{asmrpl} without the replay protection does not consume more energy than \gls{rpl} in \acrshort{umrpl}. It has been confirmed that enabling the replay protection mechanism of \gls{rpl} reduces the effect of the Neighbor attack at the expense of consuming more energy.

We proposed two suggestions to be considered when designing \gls{rpl}-based \gls{iot} networks: (i) having more routes toward the root node and (ii) reducing the "dead parent" timeouts. Evaluating each of these suggestions showed improved performance of \gls{rpl} under the investigated attacks. We argue that further investigation should be conducted on implementing both of the suggestions at the same time while having a dynamic approach for the "dead parent" timeouts optimization.

\section*{Acknowledgment}
The second and the third authors acknowledge support from the Natural Sciences and Engineering Research Council of Canada (NSERC) through the Discovery Grant program.
\bibliographystyle{IEEEtran}
\bibliography{ARaoof-Refs}
\vskip -2\baselineskip plus -1fil
\begin{IEEEbiography}[{\includegraphics[width=1in,height=1.25in,clip,keepaspectratio]{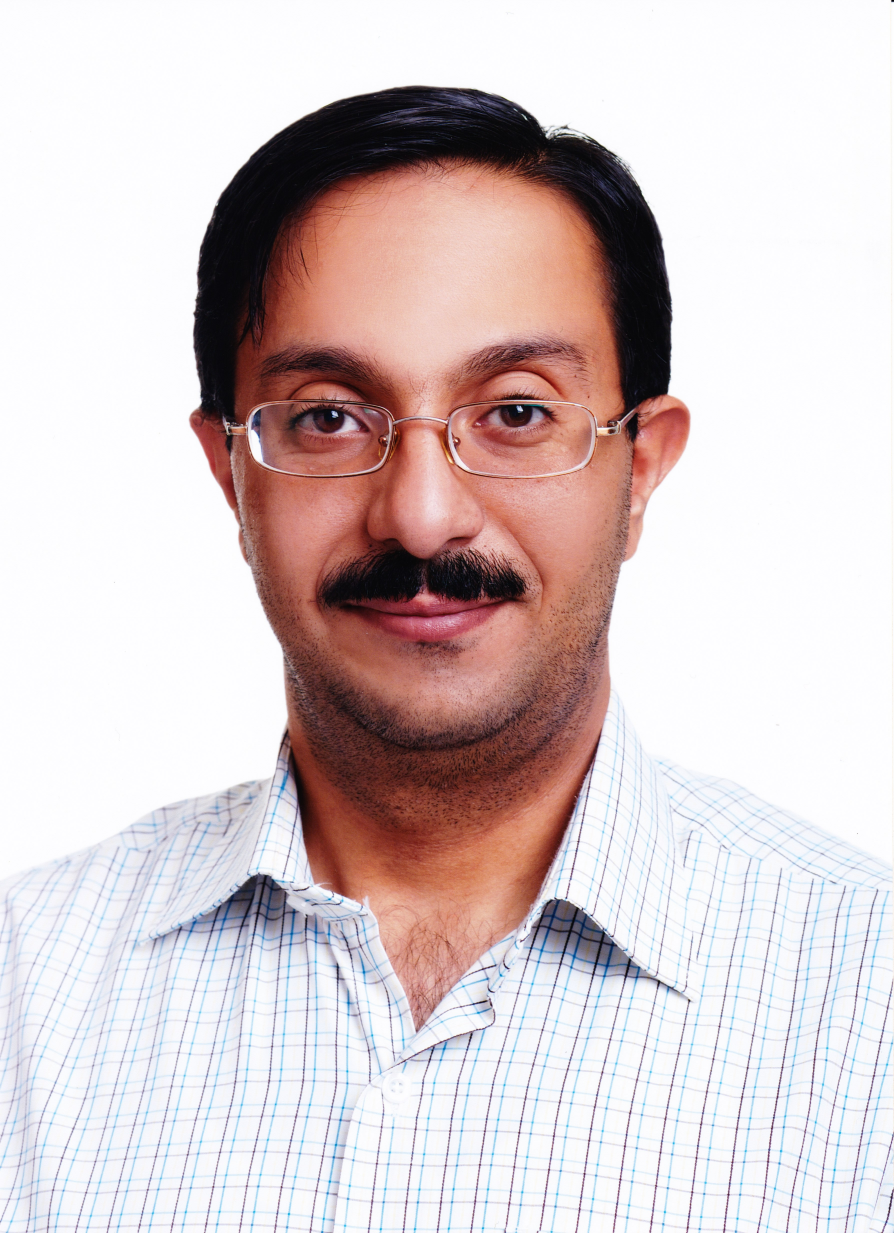}}]{Ahmed Raoof}
	is a graduate from University of Benghazi, Libya; where he received both of his degrees: a B.Sc. in Electrical and Electronic Eng. (2003), and a M.Sc. in Telecomm. Eng. (2009). After that, he worked in the same university as a lecturer at the Department of Comp. Net., Faculty of Information Technology (2009 - 2013). Currently, he is pursuing the Ph.D. degree with the Department of Sys. and Comp. Eng. at Carleton University. His research focuses on network protocols and the security of computer networks and Internet of Things.
\end{IEEEbiography}
\vskip -2\baselineskip plus -1fil
\begin{IEEEbiography}[{\includegraphics[width=1in,height=1.25in,clip,keepaspectratio]{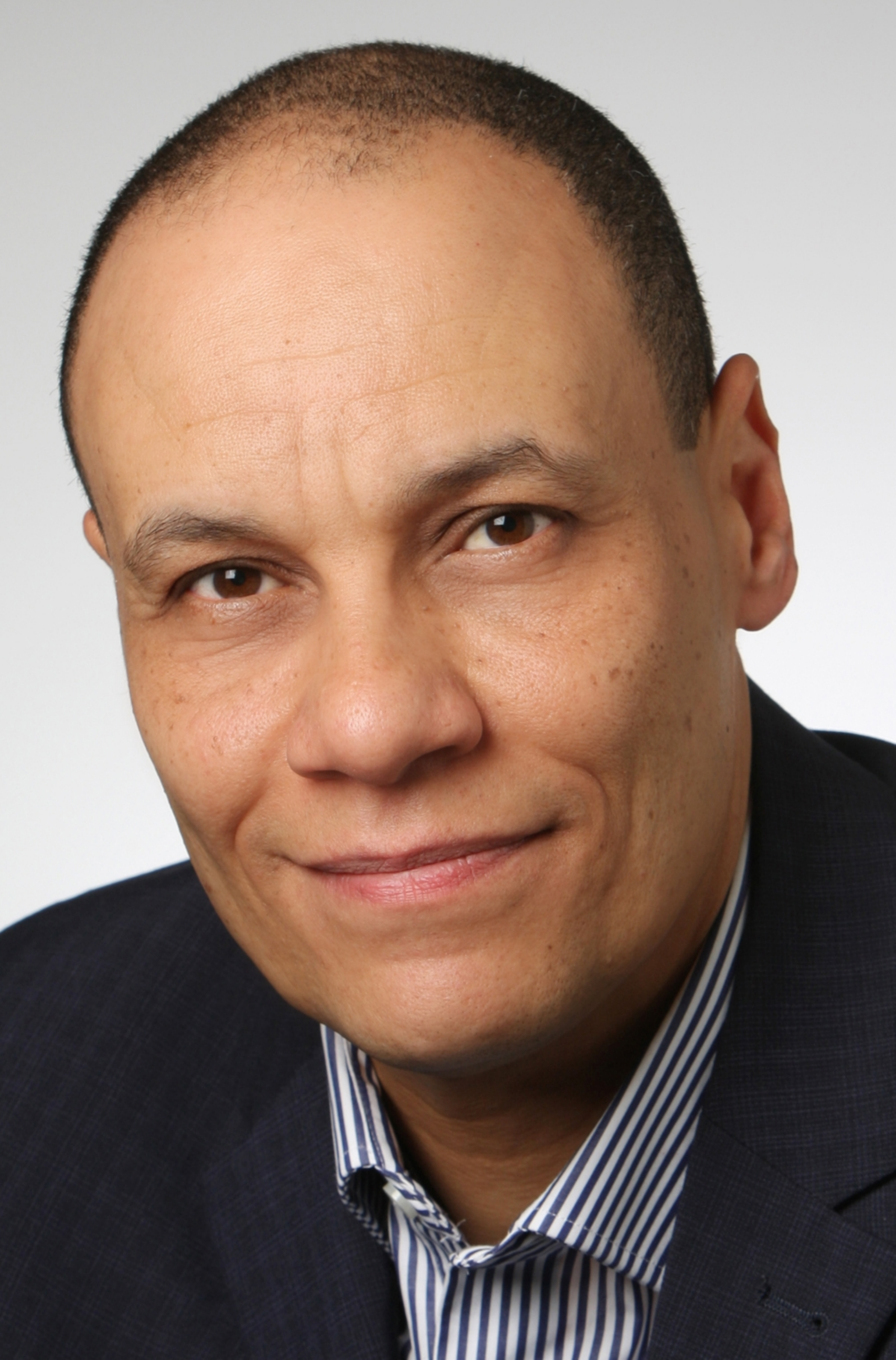}}]{Ashraf Matrawy}
	is a Professor at the School of Information Technology, Carleton University. He is a senior member of the IEEE and serves on the editorial board of the IEEE Comm. Surveys and Tutorials journal. He has served as a technical program committee member of several IEEE conferences (CNS, ICC, Globecom, LCN) and IEEE/ACM CCGRID. He is also a Network co-Investigator of Smart Cybersecurity Network (SERENE-RISC). His research interests include reliable and secure computer networks, SDN, and cloud computing. 
\end{IEEEbiography}
\vskip -2\baselineskip plus -1fil
\begin{IEEEbiography}[{\includegraphics[width=1in,height=1.25in,clip,keepaspectratio]{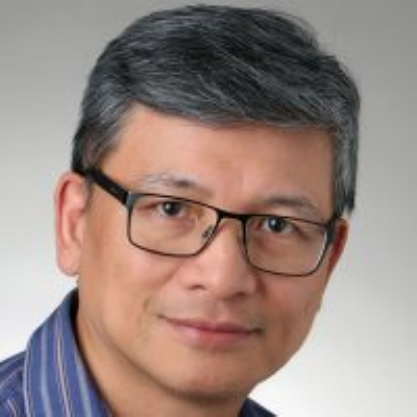}}]{Chung-Horng Lung}
	Chung-Horng Lung received the B.S. degree in Computer Science and Eng. from Chung-Yuan Christian University, Taiwan and the M.S. and Ph.D. degrees in Computer Science and Eng. from Arizona State University. He was with Nortel Networks from 1995 to 2001. In September 2001, he joined the Department of Systems and Computer Eng., Carleton University, Ottawa, Canada, where he is now a Professor. His research interests include: Communication Networks, Software Engineering, and Cloud Computing.
\end{IEEEbiography}
\vskip -2\baselineskip plus -1fil


\vfill
\end{document}